\documentclass[epj,a4paper]{svjour}
\usepackage{graphicx}
\usepackage{amsfonts}
\usepackage{amsmath}
\usepackage{hyperref}
\usepackage{float}

\begin{document}

\title{Derivation of fluid dynamics from kinetic theory with the 14--moment
approximation}
\author{G.\ S.\ Denicol\inst{1}, E.\ Moln\'ar\inst{2,3,4}, H.\ Niemi%
\inst{2,5}, and D.\ H.\ Rischke\inst{1,2}}
\institute{Institute for Theoretical Physics, Goethe University, Max-von-Laue Str. 1,
D-60438 Frankfurt am Main, Germany 
\and 
Frankfurt Institute for Advanced Studies, Ruth-Moufang-Str.\ 1, D-60438 Frankfurt am Main, Germany 
\and 
MTA Wigner Research Centre for Physics, H-1525 Budapest, P.O.Box 49, Hungary 
\and
MTA-DE Particle Physics Research Group, H-4010 Debrecen, P.O.Box 105, Hungary 
\and 
Department of Physics, P.O. Box 35 (YFL), FI-40014 University of Jyv\"{a}skyl\"{a}, Finland}

\PACS{{12.38.Mh}{} \and {24.10.Nz}{} \and {47.75.+f}{}}

\abstract{We review the traditional derivation of the fluid-dynamical equations 
from kinetic theory according to Israel and Stewart. 
We show that their procedure to close the 
fluid-dynamical equations of motion is not unique. Their
approach contains two approximations, the first being the so-called 14-moment
approximation to truncate the single-particle distribution
function. The second consists in the choice of equations of motion for
the dissipative currents. Israel and Stewart used the second moment of
the Boltzmann equation, but this is not the only possible choice. In fact, 
there are infinitely many moments of the Boltzmann equation which can
serve as equations of motion for the dissipative currents.
All resulting equations of motion have the
same form, but the transport coefficients are different in each case.
} 

\authorrunning{G.~S.~Denicol et al.}

\maketitle

\section{Introduction}

Fluid dynamics is an effective theory to describe the long-wavelength,
low-frequency dynamics of macroscopic systems. In non-relativistic systems,
the Navier-Stokes equations are able to describe a wide variety of fluids,
from weakly interacting gases, such as air, to liquids, such as water. On
the other hand, the theory of relativistic dissipative fluid dynamics has
not yet been completely established and remains a topic of intense
investigation. For dilute systems, the derivation of fluid dynamics can be
investigated starting from the relativistic Boltzmann equation.

Chapman-Enskog theory \cite{Chapman} is a well-known approach to derive
fluid-dynamical equations from the Boltzmann equation. In this approach, the
single-particle distribution function which is the solution of the Boltzmann
equation is expressed in terms of an expansion in gradients of the primary
fluid-dynamical variables, i.e., chemical potential, temperature, and
velocity, each term containing a different power or order of derivatives.
This leads to a series in powers of the Knudsen number, $\mathrm{Kn}=\lambda
/L$, the ratio of the mean-free path of the particles, $\lambda $, and a
characteristic macroscopic length, $L$. As is well-known, to zeroth order
this method leads to the equations of ideal fluid dynamics. To first order
in Knudsen number one obtains the Navier-Stokes equations of fluid dynamics.
To second and higher order in Knudsen number one obtains the Burnett and
super-Burnett equations. However, relativistic Navier-Stokes theory, as well
as any higher-order truncation of the relativistic Chapman-Enskog expansion
is unstable and, consequently, unsuitable to describe any relativistic fluid
existing in nature \cite{his,Denicol:2008ha,Pu:2009fj}.

The source of such an instability is well understood in the relativistic
case: it comes from the acausality of Navier-Stokes theory \cite%
{Denicol:2008ha,Pu:2009fj}. Therefore, a consistent and stable theory of
relativistic fluid dynamics must also be causal. Causal fluid-dynamical
equations were first derived from kinetic theory by H.\ Grad \cite{Grad},
for non-relativistic systems via the method of moments. In Grad's original
work, the single-particle distribution function is expanded around local
equilibrium in terms of a complete set of Hermite polynomials \cite%
{Grad_Hermite}. Fluid dynamics is obtained by explicitly truncating this
expansion, expressing the distribution function in terms of only 13 moments:
the velocity field, the temperature, the chemical potential, the heat
current, and the shear-stress tensor. Due to this truncation scheme the
method became known as the 13-moment approximation. In non-relativistic
systems, the correction to the equilibrium pressure, the bulk viscous
pressure, vanishes and is not included in the 13 variables.

The generalization of Grad's moment method to relativistic systems is
non-trivial. One reason is the lack of a suitable set of orthogonal
polynomials to replace the Hermite polynomials \cite%
{DeGroot,Stewart_Review,Stewart_Proc}. Despite this problem, relativistic
generalizations have been given by several authors \cite%
{Chernikov,Vignon,Marle,Kranys,Anderson}. One of the most well-known works
on this topic was done by Israel and Stewart (IS) \cite%
{Stewart_Review,Stewart_Proc,IS}. In this approach, the single-particle
distribution function is expanded in momentum space around its local
equilibrium value in terms of a series of (reducible) Lorentz-tensors formed
of particle four-momentum $k^{\mu }$, i.e., $1,k^{\mu},$ $k^{\mu }k^{\nu
},\cdots$. The procedure adopted by Israel and Stewart is very similar to
Grad's: this expansion is truncated at second order in momentum, leaving 14
moments and 14 coefficients in the distribution function to be identified,
the so-called 14-moment approximation (in the relativistic case, the bulk
viscous pressure does not vanish, leading to one additional moment when
compared to Grad's original approach). However, since the expansion is not
realized in terms of an orthogonal set, the coefficients of the truncated
expansion cannot be immediately determined. For this reason, Israel and
Stewart chose a set of constraints to express the expansion coefficients in
terms of the main fluid-dynamical variables. Furthermore, since the zeroth
and first moments of the Boltzmann equation are the usual conservation laws,
it seemed natural to choose the next (the second) moment of the Boltzmann
equation to augment and close the conservation equations.

Nevertheless, this choice is ambiguous, since, once the 14-moment
approximation is applied, any moment of the Boltzmann equation will lead to
a closed set of equations \cite{Denicol:2010xn,Denicol:2012cn}. Therefore,
inconsistencies may arise because of an ambiguity in the choice of the
moment equation for closure. Recently, it was confirmed that, at least for
some cases, the IS equations are not in good agreement with the numerical
solution of the Boltzmann equation \cite%
{Huovinen:2008te,Molnar:2009pq,El:2009vj,Bouras:2009nn,Bouras:2010nt,Bouras:2010hm}%
. Also, the transport coefficients obtained by Israel and Stewart do not
coincide with quantum-field theoretical calculations \cite{Denicol:2010br}.

In this paper, we review the derivation of the fluid-dynamical equations
from the Boltzmann equation using the 14-moment approximation, but from a
different perspective. First, the single-particle distribution function is
expanded around equilibrium in terms of an orthogonal basis. This allows us
to determine the coefficients of the expansion without assuming additional
constraints. Then, we show how the 14-moment approximation emerges from such
a complete moment expansion. Second, we obtain the fluid-dynamical equations
for an \textit{arbitrary} moment of the Boltzmann equation and discuss the
ambiguity of the 14-moment approximation. We explicitly show that the form
of the equations is always the same, regardless of the choice of moment of
the Boltzmann equation, but the values of the transport coefficients are
different. Thus, we explicitly demonstrate that the traditional 14-moment
approximation applied to the Boltzmann equation is not able to provide a
unique set of fluid-dynamical equations of motion. In Ref.~\cite%
{Denicol:2012cn} it is demonstrated how one can resolve this ambiguity by
including higher moments of the single-particle distribution function, but
this is not the subject of this paper.

The ambiguity in the 14-moment approximation is explicitly demonstrated by
calculating the transport coefficients in the 14-moment approximation for a
classical gas of massless particles with a constant cross section. We use
two different sets of moments to close the equations of motion: the one used
by Israel and Stewart \cite{IS}, and the one used in Ref.~\cite%
{Denicol:2010xn}, and show how they lead to different transport
coefficients. Note that, although our final equations contain terms that
were neglected in the papers by Israel and Stewart, see also Refs.\ \cite%
{Betz:2008me,Betz:2010cx}, we will still refer to them as IS equations.

This paper is organized as follows. In Sec.~\ref{Fluid_Dynamics} we briefly
introduce relativistic fluid dynamics and its dynamical variables. The
Boltzmann equation and the definitions of the fluid-dynamical variables from
the perspective of kinetic theory are introduced in Sec.~\ref{BoltzmannEq}.
The orthonormal basis for the moment expansion and the exact equations for
the moments are derived in Sec.~\ref{General_eqs}. In Sec.~\ref{14M} the
14-moment approximation is applied and the fluid-dynamical equations are
derived. The choice of the moment is analysed in Sec.~\ref{choice_of_moment}%
. Finally, we conclude in Sec.~\ref{conclusions}.

Throughout this work we use natural units $\hbar =k_{B}=c=1$; the metric
tensor is $g^{\mu \nu }=\text{diag}(+,-,-,-)$.

\section{Relativistic fluid dynamics}

\label{Fluid_Dynamics}

In relativistic fluid dynamics, the variables that specify the macroscopic
state of a system are the energy-momentum tensor, $T^{\mu \nu }$, and the
particle or net-charge four-current, $N^{\mu }$. Here we restrict ourselves
to only one conserved particle species or net charge. Thus, particle number
and energy-momentum conservation imply that 
\begin{eqnarray}
\partial _{\mu }N^{\mu } &=&0,  \label{bla1} \\
\partial _{\mu }T^{\mu \nu } &=&0.  \label{bla2}
\end{eqnarray}%
In relativistic fluid dynamics, it is useful to define a time-like
four-vector, $u^{\mu }(t,\mathbf{x})$, normalized to $u^{\mu }u_{\mu }=1$,
and a projection operator orthogonal to it, 
\begin{equation}
\Delta ^{\mu \nu }=g^{\mu \nu }-u^{\mu }u^{\nu },
\label{projection_operator}
\end{equation}%
where $\Delta ^{\mu \nu }u_{\nu }\equiv \Delta ^{\mu \nu }u_{\mu }=0$ and $%
\Delta _{\mu }^{\mu }=3$. Later on, $u^{\mu }$ will be identified as the
fluid four-velocity. From now on, we denote the projection orthogonal to $%
u^{\mu }$ as $A^{\left\langle \mu \right\rangle }=\Delta ^{\mu \nu }A_{\nu }$%
, valid for an arbitrary four-vector $A^{\mu }$. In case of second-rank
tensors, $A^{\mu \nu }$, we define the orthogonal and traceless projection
as $A^{\left\langle \mu \nu \right\rangle }=\Delta ^{\mu \nu \alpha \beta
}A_{\alpha \beta }$, where 
\begin{equation}
\Delta ^{\mu \nu \alpha \beta }=\frac{1}{2}\left( \Delta ^{\mu \alpha
}\Delta ^{\beta \nu }+\Delta ^{\nu \alpha }\Delta ^{\beta \mu }\right) -%
\frac{1}{3}\Delta ^{\mu \nu }\Delta ^{\alpha \beta }.
\end{equation}%
Using the projection operator from Eq. (\ref{projection_operator}), the
space-time derivative can be decomposed as 
\begin{equation}
\partial _{\mu }=u_{\mu }D+\nabla _{\mu },
\end{equation}%
where the comoving time derivative is $D=u^{\mu }\partial _{\mu },$ while
the space-like gradient is $\nabla _{\mu }=\Delta _{\mu }^{\nu }\partial
_{\nu }$. For the time derivative we also use the notation $DA=\dot{A}$.
Applying the above notation, the relativistic Cauchy-Stokes decomposition is 
\begin{equation}
\partial _{\mu }u_{\nu }=u_{\mu }\dot{u}_{\nu }+\frac{1}{3}\theta \Delta
_{\mu \nu }+\sigma _{\mu \nu }+\omega _{\mu \nu },
\end{equation}%
\newline
where the expansion scalar, $\theta $, the shear tensor, $\sigma ^{\mu \nu }$%
, and the vorticity, $\omega ^{\mu \nu }$, are defined as 
\begin{eqnarray}
\theta  &=&\nabla _{\mu }u^{\mu }, \\
\sigma ^{\mu \nu } &\equiv &\nabla ^{\langle \mu }u^{\nu \rangle }=\frac{1}{2%
}\left( \nabla ^{\mu }u^{\nu }+\nabla ^{\nu }u^{\mu }\right) -\frac{1}{3}%
\theta \Delta ^{\mu \nu }, \\
\omega ^{\mu \nu } &\equiv &\nabla ^{\lbrack \mu }u^{\nu ]}=\frac{1}{2}%
\left( \nabla ^{\mu }u^{\nu }-\nabla ^{\nu }u^{\mu }\right) .
\end{eqnarray}%
The particle four-current and energy-momentum tensor can be decomposed with
respect to $u^{\mu }$ as 
\begin{eqnarray}
N^{\mu } &=&nu^{\mu }+V^{\mu },  \label{N_mu} \\
T^{\mu \nu } &=&\varepsilon u^{\mu }u^{\nu }-P\Delta ^{\mu \nu }+u^{\nu
}W^{\mu }+u^{\mu }W^{\nu }+\pi ^{\mu \nu },  \label{T_mu_nu}
\end{eqnarray}%
where $n=N^{\mu }u_{\mu }$ is the particle density and $\varepsilon =u_{\mu
}T^{\mu \nu }u_{\nu }$ is the energy density. The trace of the
energy-momentum tensor, $P=-\frac{1}{3}\Delta _{\mu \nu }T^{\mu \nu }$,
denotes the isotropic pressure. The latter is defined as the sum of the
equilibrium pressure $P_{0}$, and the bulk viscous pressure $\Pi $, $%
P=P_{0}+\Pi $. The particle diffusion current is defined as 
\begin{equation}
V^{\mu }=\Delta _{\nu }^{\mu }N^{\nu },
\end{equation}%
while the energy-momentum diffusion current is 
\begin{equation}
W^{\mu }=\Delta ^{\mu \alpha }T_{\alpha \beta }u^{\beta }.
\end{equation}%
The shear-stress tensor, $\pi ^{\mu \nu }$, is that part of the
energy-momentum tensor that is symmetric, traceless, and orthogonal to $%
u^{\mu }$, 
\begin{equation}
\pi ^{\mu \nu }=T^{\langle \mu \nu \rangle }.
\end{equation}%
In local thermal equilibrium, the decompositions of Eqs. (\ref{N_mu},\ref%
{T_mu_nu}) reduce to the ideal-fluid form 
\begin{eqnarray}
N_{0}^{\mu } &=&n_{0}u^{\mu },  \label{N0_mu} \\
T_{0}^{\mu \nu } &=&\varepsilon _{0}u^{\mu }u^{\nu }-P_{0}\Delta ^{\mu \nu }.
\label{T0_mu_nu}
\end{eqnarray}%
Local thermodynamic equilibrium guarantees that the particle density $n_{0}$, 
entropy density $s_{0}$, energy density $\varepsilon _{0}$, and
thermodynamic pressure $P_{0}$, are related to the temperature, $T$, and
chemical potential, $\mu $, through an equation of state (EoS), i.e., $%
P_{0}=P_{0}(T,\mu )$, from which one can obtain 
\begin{equation}
n_{0}=\frac{\partial P_{0}}{\partial \mu }\;,\;\;\;s_{0}=\frac{\partial P_{0}%
}{\partial T}\;,\ 
\end{equation}%
and 
\begin{equation}
\varepsilon _{0}=Ts_{0}-P_{0}+\mu n_{0}\;.
\end{equation}

In general, the choice of $u^{\mu }$ is ambiguous. The frame where $u^{\mu
}\equiv u_{LR}^{\mu }=(1,0,0,0)$ is called the local rest frame (LRF) of
matter. From the physical perspective, there are two natural choices which
fix the LRF but at the same time promote $u^{\mu }$ to a dynamical quantity.
According to the definition of Landau and Lifshitz \cite{Landau}, the LRF is
tied to the flow of energy-momentum, which leads to 
\begin{equation}
u^{\mu }=\frac{T^{\mu \nu }u_{\nu }}{\sqrt{T^{\mu \alpha }u_{\alpha }T_{\mu
\beta }u^{\beta }}}\;,  \label{Landau_flow}
\end{equation}%
and thus the energy-momentum diffusion current vanishes, $W^{\mu }=0$.

The choice of Eckart \cite{Eckart} relates the LRF to the flow of conserved
particles as 
\begin{equation}
u^{\mu }=\frac{N^{\mu }}{\sqrt{N^{\nu }N_{\nu }}}\;,  \label{Eckart_flow}
\end{equation}%
which implies that the diffusion current vanishes, $V^{\mu }=0$. Sometimes
it is convenient to introduce the heat flow, 
\begin{equation}  \label{heatflow}
q^{\mu }=W^{\mu }-h\; V^{\mu }\;,
\end{equation}
where $h = (\varepsilon +P_0)/n$ is the enthalpy per particle (or per net
charge).

Once the four-flow of matter is specified, i.e., replacing the three
independent components of $W^{\mu }$ or $V^{\mu }$ by $u^{\mu }$, we still
have to determine 15 independent dynamical variables: six variables, $u^{\mu
}$, $\varepsilon $, $n $, and $P_{0}$, as in the case of an ideal fluid, and
nine variables related to dissipation, $\Pi $, $q^{\mu }$, and $\pi ^{\mu
\nu }$. Note that the EoS gives one additional constraint and therefore
reduces the number of independent variables to 14.

The conservation laws (\ref{bla1},\ref{bla2}) constitute only five
equations. Thus, to properly close the fluid-dynamical equations it is
necessary to introduce nine additional equations which determine the
evolution of the remaining dissipative fields, $\Pi $, $q^{\mu }$, and $\pi
^{\mu \nu } $. The relativistic extension of Navier-Stokes theory relates
the dissipative quantities to gradients of the primary fluid-dynamical
fields, 
\begin{eqnarray}
\Pi &=&-\zeta \theta , \\
q^{\mu } &=&-\kappa _{q}\frac{T^{2}}{h}\nabla ^{\mu }\left( \frac{\mu }{T}%
\right) , \\
\pi ^{\mu \nu } &=&2\eta \sigma ^{\mu \nu },
\end{eqnarray}%
where the bulk viscosity coefficient $\zeta $, the heat-flow coefficient $%
\kappa _{q}$, and the shear viscosity coefficient $\eta $ are
positive-definite functions of $T$ and $\mu$.

However, as mentioned in the introduction this naive approach leads to
intrinsic problems, such as acausal signal propagation and instabilities,
and is therefore unsuitable to describe relativistic fluids. The acausality
problems were solved by introducing memory effects into the definitions of $%
\Pi $, $q^{\mu }$, and $\pi ^{\mu \nu }$, which are no longer assumed to be
linearly related to gradients of the primary fluid-dynamical variables \cite%
{Israel:1976tn,Koide:2006ef,Denicol:2009zz,Liu86,Geroch:1990bw,Geroch:1995bx,Muller:1999in}%
. Instead, they become independent dynamical variables that obey dynamical
equations of motion (which introduce the relaxation times $\tau _{\pi }$, $%
\tau _{\Pi }$, and $\tau _{q}$) that describe their transient dynamics
towards their respective asymptotic relativistic Navier-Stokes solution, 
\begin{eqnarray}
\tau _{\Pi }\,\dot{\Pi}+\Pi &=&-\zeta \theta +\ldots , \\
\tau _{q}\,\dot{q}^{\left\langle \mu \right\rangle }+n^{\mu } &=&-\kappa _{q}%
\frac{T^{2}}{h}\nabla ^{\mu }\left( \frac{\mu }{T}\right) +\ldots , \\
\tau _{\pi }\,\dot{\pi}^{\left\langle \mu \nu \right\rangle }+\pi ^{\mu \nu
} &=&2\eta \,\sigma ^{\mu \nu }+\ldots ,
\end{eqnarray}%
where $\dot{q}^{\left\langle \mu \right\rangle }=\Delta _{\alpha }^{\mu
}Dq^{\alpha }$ and $\dot{\pi}^{\left\langle \mu \nu \right\rangle }=\Delta
_{\alpha \beta }^{\mu \nu }D\pi ^{\alpha \beta }$ and the dots denote
possible higher-order terms. These are the type of equations of motion which
can also be derived from relativistic kinetic theory as shown by Israel and
Stewart and others \cite{Stewart_Review,IS,Betz:2010cx,Muronga:2006zx}.
Causality is guaranteed, provided the relaxation times fulfill certain
constraints \cite{Pu:2009fj}.

\section{The relativistic Boltzmann equation}

\label{BoltzmannEq}

Let us consider a relativistic dilute gas characterized only by the
single-particle distribution function $f_{\mathbf{k}}\equiv f(x^{\mu
},k^{\mu })$, the evolution of which is given by the relativistic Boltzmann
equation \cite{DeGroot}, 
\begin{equation}
k^{\mu }\partial _{\mu }f_{\mathbf{k}}=C\left[ f\right] ,
\label{Boltzmann_Eq}
\end{equation}%
where $k^{\mu }=(k^{0},\mathbf{k})$ with $k^{0}=\sqrt{\mathbf{k}^{2}+m^{2}}$
and $m$ being the mass of the particles. For the collision term $C\left[ f%
\right] $, we consider only elastic two-to-two collisions with incoming
momenta $k,k^{\prime }$, and outgoing momenta $p,p^{\prime }$, 
\begin{align}
C\left[ f\right] & =\frac{1}{\nu }\int dK^{\prime }dPdP^{\prime }W_{\mathbf{%
kk}\prime \rightarrow \mathbf{pp}\prime }  \notag \\
& \times \left( f_{\mathbf{p}}f_{\mathbf{p}^{\prime }}\tilde{f}_{\mathbf{k}}%
\tilde{f}_{\mathbf{k}^{\prime }}-f_{\mathbf{k}}f_{\mathbf{k}^{\prime }}%
\tilde{f}_{\mathbf{p}}\tilde{f}_{\mathbf{p}^{\prime }}\right) ,
\label{Col_term}
\end{align}%
where $\nu =2$ is a symmetry factor. The Lorentz-invariant phase volume is $%
dK\equiv \,gd^{3}\mathbf{k/}\left[ (2\pi )^{3}k^{0}\right] ,$ with $g$ being
the number of internal degrees of freedom. The Lorentz-invariant transition
rate $W_{\mathbf{kk}\prime \rightarrow \mathbf{pp}\prime }$ is symmetric
with respect to the exchange of the outgoing momenta, as well as to time
reversal, 
\begin{equation}
W_{\mathbf{kk}\prime \rightarrow \mathbf{pp}\prime }\equiv W_{\mathbf{kk}%
\prime \rightarrow \mathbf{p}\prime \mathbf{p}}=W_{\mathbf{pp}\prime
\rightarrow \mathbf{kk}\prime }.  \label{detailed_balance}
\end{equation}%
Here, we also take into account quantum statistics and introduced the
notation $\tilde{f}_{\mathbf{k}}\equiv 1-af_{\mathbf{k}}$, where $a=1$ ($a=-1
$) for fermions (bosons) and $a=0$ in the limiting case of classical
Boltzmann-Gibbs statistics.

The particle four-flow and the energy-momentum tensor are identified as the
first and second moments of the single-particle distribution function, 
\begin{eqnarray}
N^{\mu } &=&\left\langle k^{\mu }\right\rangle ,  \label{charge} \\
T^{\mu \nu } &=&\left\langle k^{\mu }k^{\nu }\right\rangle ,  \label{Tmunu}
\end{eqnarray}%
where we adopted the following notation for the averages 
\begin{equation}
\left\langle \ldots \right\rangle =\int dK\text{ }\left( \ldots \right) \ f_{%
\mathbf{k}}.
\end{equation}%
Making use of the properties of the transition rate (\ref{detailed_balance}), one
can show \cite{DeGroot} that the particle four-flow and the energy-momentum
tensor satisfy the conservation equations (\ref{bla1},\ref{bla2}) for any
solution of the Boltzmann equation, 
\begin{eqnarray}
\partial _{\mu }\left\langle k^{\mu }\right\rangle &\equiv &\int dKC\left[ f%
\right] =0,  \label{Eq_1} \\
\partial _{\mu }\left\langle k^{\mu }k^{\nu }\right\rangle &\equiv &\int
dKk^{\nu }C\left[ f\right] =0.  \label{Eq_2}
\end{eqnarray}

In order to identify the macroscopic variables introduced in Eqs.\ (\ref%
{N_mu},\ \ref{T_mu_nu}) in terms of the single-particle distribution
function we decompose the momentum of the particles $k^{\mu }$ into two
parts: one parallel to the flow velocity $u^{\mu }$ and the other orthogonal
to the latter, 
\begin{equation}
k^{\mu }=E_{\mathbf{k}}u^{\mu }+k^{\left\langle \mu \right\rangle },
\label{k_mu_decomp}
\end{equation}%
where we defined the energy of a particle as $E_{\mathbf{k}}\equiv u_{\mu
}k^{\mu }$. Using the above decomposition in Eqs.\ (\ref{charge},\ref{Tmunu}%
) we obtain 
\begin{align}
N^{\mu }& =\left\langle E_{\mathbf{k}}\right\rangle u^{\mu }+\left\langle
k^{\left\langle \mu \right\rangle }\right\rangle ,  \label{Nk_mu} \\
T^{\mu \nu }& =\left\langle E_{\mathbf{k}}^{2}\right\rangle u^{\mu }u^{\nu }+%
\frac{1}{3}\Delta ^{\mu \nu }\left\langle \Delta ^{\alpha \beta }k_{\alpha
}k_{\beta }\right\rangle   \notag \\
& +u^{\nu }\left\langle E_{\mathbf{k}}k^{\left\langle \mu \right\rangle
}\right\rangle +u^{\mu }\left\langle E_{\mathbf{k}}k^{\left\langle \nu
\right\rangle }\right\rangle +\left\langle k^{\left\langle \mu \right.
}k^{\left. \nu \right\rangle }\right\rangle .  \label{Tk_mu_nu}
\end{align}%
Comparing these decompositions with Eqs. (\ref{N_mu},\ref{T_mu_nu}), we
identify the main fluid-dynamical quantities as averages or moments with
respect to an arbitrary solution of the Boltzmann equation, 
\begin{eqnarray}
n &=&\left\langle E_{\mathbf{k}}\right\rangle \,,\ \varepsilon =\left\langle
E_{\mathbf{k}}^{2}\right\rangle \,,\ P_{0}+\Pi =-\frac{1}{3}\left\langle
\Delta ^{\mu \nu }k_{\mu }k_{\nu }\right\rangle ,  \notag \\
V^{\mu } &=&\left\langle k^{\left\langle \mu \right\rangle }\right\rangle
\,,\;W^{\mu }=\left\langle E_{\mathbf{k}}k^{\left\langle \mu \right\rangle
}\right\rangle \,,\text{ }\pi ^{\mu \nu }=\left\langle k^{\left\langle \mu
\right. }k^{\left. \nu \right\rangle }\right\rangle \,.\ \ \ \ \ 
\label{def_hy_qua}
\end{eqnarray}%
Similarly, we introduce the average with respect to the local equilibrium
distribution function $f_{0\mathbf{k}}$, 
\begin{equation}
\langle \ldots \rangle _{0}=\int dK\left( \ldots \right) f_{0\mathbf{k}},
\end{equation}%
where 
\begin{equation}
f_{0\mathbf{k}}(x^{\mu },k^{\mu })=\left[ \exp \left( \beta _{0}E_{\mathbf{k}%
}-\alpha _{0}\right) +a\right] ^{-1}.
\end{equation}%
Although $f_{0\mathbf{k}}$ satisfies detailed balance, it is not a solution
of the Boltzmann equation. The quantities $\alpha _{0}(x^{\mu })$ and $\beta
_{0}(x^{\mu })$ are defined for an arbitrary non-equilibrium distribution
function $f_{\mathbf{k}}$ by the matching conditions, 
\begin{equation}
n\equiv n_{0}=\langle E_{\mathbf{k}}\rangle _{0},\ \varepsilon \equiv
\varepsilon _{0}=\left\langle E_{\mathbf{k}}^{2}\right\rangle _{0}.
\label{matching}
\end{equation}%
In local equilibrium we would then identify $\beta _{0}=1/T$ as the inverse
temperature and $\alpha _{0}=\mu /T$ as the ratio of chemical potential over
temperature. The matching conditions (\ref{matching}) lead to 
\begin{equation}
\langle E_{\mathbf{k}}\rangle _{\delta }=0,\ \langle E_{\mathbf{k}%
}^{2}\rangle _{\delta }=0,  \label{matching_0}
\end{equation}%
where $\left\langle \ldots \right\rangle _{\delta }\equiv \left\langle
\ldots \right\rangle -\left\langle \ldots \right\rangle _{0}$. The matching
conditions (\ref{matching}) are convenient as they allow us to use
equilibrium thermodynamic relations between $n$, $\varepsilon $, $P_{0}$, $T$%
, and $\mu $. We also note that the EoS is not an additional input, but
follows from the single-particle distribution function in local equilibrium.

Finally, the separation between thermodynamic pressure and bulk viscous
pressure is achieved as 
\begin{equation}
P_{0}=-\frac{1}{3}\,\left\langle \Delta ^{\mu \nu }k_{\mu }k_{\nu
}\right\rangle _{0},\ \Pi =-\frac{1}{3}\,\left\langle \Delta ^{\mu \nu
}k_{\mu }k_{\nu }\right\rangle _{\delta }.  \label{bulk_matching}
\end{equation}%
We remark that $\left\langle k^{\left\langle \mu \right\rangle
}\right\rangle _{0}=\left\langle E_{\mathbf{k}}k^{\left\langle \mu
\right\rangle }\right\rangle \,_{0}=\left\langle k^{\left\langle \mu \right.
}k^{\left. \nu \right\rangle }\right\rangle _{0}=0$ guarantees that in local
thermal equilibrium all dissipative currents vanish.

\section{Moment expansion}

\label{General_eqs}

The method of moments is the most common approach to derive the so-called
second-order theories from kinetic theory. In this section, we review the
basic ideas of this method along the lines of Refs.\ \cite%
{DeGroot,Stewart_Review}. Since we are interested in near-equilibrium
solutions of the Boltzmann equation, we start by expanding $f_{\mathbf{k}}$
around the local equilibrium distribution function $f_{0\mathbf{k}}$, 
\begin{equation}
f_{\mathbf{k}}\equiv f_{0\mathbf{k}}+\delta f_{\mathbf{k}}=f_{0\mathbf{k}%
}\left( 1+\tilde{f}_{0\mathbf{k}}\phi _{\mathbf{k}}\right) ,
\label{small_correc}
\end{equation}%
where $\phi _{\mathbf{k}}$ represents a general non-equilibrium correction.

Following Ref.\ \cite{DeGroot}, $\phi _{\mathbf{k}}$ is expanded in momentum
space with the help of the irreducible tensors $1$, $k^{\left\langle \mu
\right\rangle }$, $k^{\left\langle \mu \right. }k^{\left. \nu \right\rangle
} $, $k^{\left\langle \mu \right. }k^{\nu }k^{\left. \lambda \right\rangle }$%
, $\cdots $, forming a complete and orthogonal set, analogous to the
spherical harmonics \cite{Anderson}. These irreducible tensors are defined
by using the symmetrized traceless projections as 
\begin{equation}
k^{\left\langle \mu _{1}\right. } \cdots k^{\left. \mu _{m}\right\rangle }=
\Delta ^{\mu _{1} \cdots \mu _{m}\nu _{1} \cdots \nu _{m}}k_{\nu _{1}}
\cdots k_{\nu _{m}},
\end{equation}%
see also Ref.\ \cite{struchtrup}. The tensors $k^{\left\langle \mu
_{1}\right. } \cdots k^{\left. \mu _{m}\right\rangle }$ satisfy the
following orthogonality condition, 
\begin{eqnarray}
&&\int dKF_{\mathbf{k}}k^{\left\langle \mu _{1}\right. } \cdots k^{\left.
\mu _{m}\right\rangle }k^{\left\langle \nu _{1}\right. } \cdots k^{\left.
\nu _{n}\right\rangle }  \notag \\
&=&\frac{m!\delta _{mn}}{\left( 2m+1\right) !!}\Delta ^{\mu _{1}\cdots \mu
_{m}\nu _{1} \cdots \nu _{m}}\int dKF_{\mathbf{k}}\left( \Delta ^{\alpha
\beta }k_{\alpha }k_{\beta }\right) ^{m}. \quad \   \label{orthogonality1}
\end{eqnarray}%
Here $m,n=0,1,2,\ldots $, $F_{\mathbf{k}}$ is an arbitrary scalar function
of $E_{\mathbf{k}}$, and $\left( 2m+1\right) !!$ denotes the double
factorial.

Using these tensors as the basis of the expansion, the non-equilibrium
correction can be written as, 
\begin{equation}
\phi _{\mathbf{k}}=\sum_{\ell =0}^{\infty }\lambda _{\mathbf{k}%
}^{\left\langle \mu _{1}\cdots \mu _{\ell }\right\rangle }\,k_{\left\langle
\mu _{1}\right. }\cdots k_{\left. \mu _{\ell }\right\rangle },
\label{expansion1}
\end{equation}%
where the index $\ell $ indicates the rank of the tensor $\lambda _{\mathbf{k%
}}^{\left\langle \mu _{1}\cdots \mu _{\ell }\right\rangle }$, and $\ell =0$
corresponds to the scalar $\lambda $. The coefficients $\lambda _{\mathbf{k}%
}^{\left\langle \mu _{1}\cdots \mu _{\ell }\right\rangle }$ may be further
expanded in energy $E_{\mathbf{k}}$ with another orthogonal basis of
functions $P_{\mathbf{k}n}^{\left( \ell \right) }$, 
\begin{equation}
\lambda _{\mathbf{k}}^{\left\langle \mu _{1}\cdots \mu _{\ell }\right\rangle
}=\sum_{n=0}^{N_{\ell }}c_{n}^{\left\langle \mu _{1}\cdots \mu _{\ell
}\right\rangle }P_{\mathbf{k}n}^{\left( \ell \right) },  \label{expansion2}
\end{equation}%
where $c_{n}^{\left\langle \mu _{1}\cdots \mu _{\ell }\right\rangle }$ are
as of yet undetermined coefficients and $N_{\ell }$ is the number of
functions $P_{\mathbf{k}n}^{\left( \ell \right) } $ considered to describe
the $\ell $-th rank tensor $\lambda _{\mathbf{k}}^{\left\langle \mu
_{1}\cdots \mu _{\ell }\right\rangle }$. The functions $P_{\mathbf{k}%
n}^{\left( \ell \right) }$ are chosen to be polynomials of order $n$ in
energy, 
\begin{equation}
P_{\mathbf{k}n}^{\left( \ell \right) }=\sum_{r=0}^{n}a_{nr}^{(\ell )}E_{%
\mathbf{k}}^{r},  \label{Poly}
\end{equation}%
and are constructed using the following orthonormality condition, 
\begin{equation}
\int dK\,\omega ^{\left( \ell \right) }\,P_{\mathbf{k}m}^{\left( \ell
\right) }P_{\mathbf{k}n}^{\left( \ell \right) }=\delta _{mn},
\label{conditions}
\end{equation}%
where the weight $\omega ^{\left( \ell \right) }$ is defined as%
\begin{equation}
\omega ^{\left( \ell \right) }= \frac{\mathcal{N}^{\left( \ell \right) }}{%
\left( 2\ell +1\right) !!}\left( \Delta ^{\alpha \beta }k_{\alpha }k_{\beta
}\right) ^{\ell }f_{0\mathbf{k}}\tilde{f}_{0\mathbf{k}}\;.
\end{equation}%
The coefficients $a_{nr}^{(\ell )}$ and the normalization constants $%
\mathcal{N}^{\left( \ell \right) }$ can be found via Gram-Schmidt
orthogonalization using the orthonormality condition (\ref{conditions}), see
Sec.\ \ref{14M} or Ref.\ \cite{Denicol:2012cn} for more details.

Finally, the single-particle distribution function can be expressed as 
\begin{equation}
f_{\mathbf{k}}=f_{0\mathbf{k}}\left( 1+\tilde{f}_{0\mathbf{k}}\sum_{\ell
=0}^{\infty }\sum_{n=0}^{N_{\ell }}\mathcal{H}_{\mathbf{k}n}^{\left( \ell
\right) }\rho _{n}^{\mu _{1}\cdots \mu _{\ell }}k_{\left\langle \mu
_{1}\right. }\cdots k_{\left. \mu _{\ell }\right\rangle }\right) ,\ 
\label{main_f_expansion}
\end{equation}%
where we introduced the energy-dependent coefficients 
\begin{equation}
\mathcal{H}_{\mathbf{k}n}^{\left( \ell \right) }=\frac{\mathcal{N}^{\left(
\ell \right) }}{\ell !}\sum_{m=n}^{N_{\ell }}a_{mn}^{(\ell )}P_{\mathbf{k}%
m}^{\left( \ell \right) }  \label{Hk}
\end{equation}%
and the generalized irreducible moment of $\delta f_{\mathbf{k}}$, 
\begin{equation}
\rho _{r}^{\mu _{1}\cdots \mu _{\ell }}= \left\langle E_{\mathbf{k}}^{r}%
\text{ }k^{\left\langle \mu _{1}\right. }\cdots k^{\left. \mu _{\ell
}\right\rangle }\right\rangle _{\delta },  \label{rho}
\end{equation}%
with 
\begin{equation}
\left\langle \ldots \right\rangle _{\delta }=\int dK\left( \ldots \right)
\delta f_{\mathbf{k}}.
\end{equation}

Using this notation, the expansion coefficients in Eq. (\ref{expansion2})
can be immediately determined using Eqs. (\ref{orthogonality1}) and (\ref%
{conditions}). For $n\leq N_{\ell }$ they are given by 
\begin{align}
c_{n}^{\left\langle \mu _{1}\cdots \mu _{\ell }\right\rangle }& \equiv \frac{%
\mathcal{N}^{\left( \ell \right) }}{\ell !}\left\langle P_{\mathbf{k}%
n}^{\left( \ell \right) }\text{ }k^{\left\langle \mu _{1}\right. }\cdots
k^{\left. \mu _{\ell }\right\rangle }\right\rangle _{\delta }  \notag \\
& =\frac{\mathcal{N}^{\left( \ell \right) }}{\ell !}\sum_{r=0}^{n}\rho
_{r}^{\mu _{1}\cdots \mu _{\ell }}a_{nr}^{(\ell )}.  \label{c_coeff}
\end{align}

Naturally, the dissipative currents are related to the tensors $\rho
_{r}^{\mu _{1}\cdots \mu _{\ell }}$. According to Eqs.\ (\ref{def_hy_qua})
we can identify them as 
\begin{eqnarray}
\rho _{0} &=&-3\Pi /m^{2},  \label{rho_0} \\
\rho _{0}^{\mu } &=&V^{\mu },  \label{rho_mu_0} \\
\rho _{1}^{\mu } &=&W^{\mu },  \label{rho_mu_1} \\
\rho _{0}^{\mu \nu } &=&\pi ^{\mu \nu }.  \label{rho_mu_nu_0}
\end{eqnarray}%
Furthermore, the matching conditions imposed in Eq.\ (\ref{matching_0}) can
also be recast using the irreducible moments, 
\begin{equation}
\rho _{1}=\rho _{2}=0.  \label{matching_rho_0_1}
\end{equation}%
The definition of the LRF corresponding to Landau's choice (\ref{Landau_flow}%
) requires that 
\begin{equation}
\rho _{1}^{\mu }=0,
\end{equation}%
while Eckart's definition (\ref{Eckart_flow}) leads to 
\begin{equation}
\rho _{0}^{\mu }=0.
\end{equation}

\subsection{General equations of motion}

\label{EoM_Gen}

So far, the single-particle distribution function was expressed in terms of
the irreducible tensors $\rho _{n}^{\mu _{1}\cdots \mu _{\ell }}$. The
time-evolution equations for these tensors can be obtained directly from the
Boltzmann equation by applying the comoving derivative to Eq.\ (\ref{rho})
together with the symmetrized traceless projection, 
\begin{equation}
\dot{\rho}_{r}^{\left\langle \mu _{1}\cdots \mu _{\ell }\right\rangle
}=\Delta _{\nu _{1}\cdots \nu _{\ell }}^{\mu _{1}\cdots \mu _{\ell }}\frac{d%
}{d\tau }\int dKE_{\mathbf{k}}^{r}k^{\left\langle \nu _{1}\right. }\cdots
k^{\left. \nu _{\ell }\right\rangle }\delta f_{\mathbf{k}}.
\label{time_deriv}
\end{equation}
Now, using the Boltzmann equation (\ref{Boltzmann_Eq}) in the form 
\begin{align}
\delta \dot{f}_{\mathbf{k}}& =-\dot{f}_{0\mathbf{k}}-E_{\mathbf{k}%
}^{-1}k_{\nu }\nabla ^{\nu }f_{0\mathbf{k}}-E_{\mathbf{k}}^{-1}k_{\nu
}\nabla ^{\nu }\delta f_{\mathbf{k}}  \notag \\
& +E_{\mathbf{k}}^{-1}C\left[ f\right] \;,  \label{linBoltz}
\end{align}%
and substituting into Eq.\ (\ref{time_deriv}), we obtain the \textit{exact}
equations for $\rho_{r}^{ \mu _{1}\cdots \mu _{\ell}}$.

Since fluid dynamics does not involve tensors of rank higher than two, it is
sufficient to derive the time-evolution equations for the fields $\rho _{r}$%
, $\rho _{r}^{\mu }$, and $\rho _{r}^{\mu \nu }$ only. Similar equations
could also be derived for the higher-rank irreducible tensors, if needed.
Thus, using Eqs.\ (\ref{time_deriv}) and (\ref{linBoltz}), the equation for
an arbitrary scalar moment is 
\begin{align}
\dot{\rho}_{r}& =C_{r-1}+\alpha _{r}^{\left( 0\right) }\theta + \left(r\rho
_{r-1}^{\mu } +\frac{G_{2r}}{D_{20}}W^{\mu }\right)\dot{u}_{\mu }  \notag \\
& -\nabla _{\mu }\rho _{r-1}^{\mu }+\frac{G_{3r}}{D_{20}}\partial _{\mu
}V^{\mu }-\frac{G_{2r}}{D_{20}}\partial _{\mu }W^{\mu }  \notag \\
& +\frac{1}{3}\left[ \left( r-1\right) m^{2}\rho _{r-2}-\left( r+2\right)
\rho _{r}-3\frac{G_{2r}}{D_{20}}\Pi \right] \theta  \notag \\
& +\left[ \left( r-1\right) \rho _{r-2}^{\mu \nu }+\frac{G_{2r}}{D_{20}}\pi
^{\mu \nu }\right] \sigma _{\mu \nu }.  \label{scalar_rho}
\end{align}%
Similarly, the time-evolution equation for the vector moment is 
\begin{align}
\dot{\rho}_{r}^{\left\langle \mu \right\rangle }& =C_{r-1}^{\left\langle \mu
\right\rangle }+\alpha _{r}^{\left( 1\right) }\nabla ^{\mu }\alpha
_{0}-\alpha _{r}^{h}\dot{W}^{\mu }+r\rho _{r-1}^{\mu \nu }\dot{u}_{\nu } 
\notag \\
& +\frac{1}{3}\left[r m^{2}\rho _{r-1}-\left( r+3\right) \rho _{r+1}-3\alpha
_{r}^{h}\Pi \right] \dot{u}^{\mu }  \notag \\
& -\frac{1}{3}\nabla ^{\mu }\left( m^{2}\rho _{r-1}-\rho _{r+1}\right)
+\alpha _{r}^{h}\nabla ^{\mu }\Pi  \notag \\
& -\Delta _{\nu }^{\mu }\left( \nabla _{\lambda }\rho _{r-1}^{\nu \lambda
}+\alpha _{r}^{h}\partial _{\lambda }\pi ^{\nu \lambda }\right)  \notag \\
& +\frac{1}{3}\left[ \left( r-1\right) m^{2}\rho _{r-2}^{\mu }-\left(
r+3\right) \rho _{r}^{\mu }-4\alpha _{r}^{h}W^{\mu }\right] \theta  \notag \\
& +\frac{1}{5}\left[ \left( 2r-2\right) m^{2}\rho _{r-2}^{\nu }-\left(
2r+3\right) \rho _{r}^{\nu }-5\alpha _{r}^{h}W^{\nu }\right] \sigma _{\nu
}^{\mu }  \notag \\
& +\left( \rho _{r}^{\nu }+\alpha _{r}^{h}W^{\nu }\right) \omega _{\left.{\,}%
\right.\nu }^{\mu } +\left( r-1\right) \rho _{r-2}^{\mu \nu \lambda }\sigma
_{\nu \lambda },  \label{vector_rho}
\end{align}%
while the equation for $\rho _{r}^{\mu \nu }$ is 
\begin{align}
\dot{\rho}_{r}^{\left\langle \mu \nu \right\rangle }& =C_{r-1}^{\left\langle
\mu \nu \right\rangle }+2\alpha _{r}^{\left( 2\right) }\sigma ^{\mu \nu }+%
\frac{2}{15}\left[ \left( r-1\right) m^{4}\rho _{r-2}\right.  \notag \\
& \left. -\left( 2r+3\right) m^{2}\rho _{r}+(r+4)\rho _{r+2}\right] \sigma
^{\mu \nu }  \notag \\
& +\frac{2}{5}\left[ r m^{2}\rho _{r-1}^{\left\langle \mu \right. }-\left(
r+5\right) \rho _{r+1}^{\left\langle \mu \right. }\right] \dot{u}^{\left.
\nu \right\rangle }  \notag \\
& +r\rho _{r-1}^{\mu \nu \lambda }\dot{u}_{\lambda }-\frac{2}{5}\nabla
^{\left\langle \mu \right. }\left( m^{2}\rho _{r-1}^{\left. \nu
\right\rangle }-\rho _{r+1}^{\left. \nu \right\rangle }\right)  \notag \\
& +\frac{1}{3}\left[ \left( r-1\right) m^{2}\rho _{r-2}^{\mu \nu }-\left(
r+4\right) \rho _{r}^{\mu \nu }\right] \theta  \notag \\
& +\frac{2}{7}\left[ \left( 2r-2\right) m^{2}\rho _{r-2}^{\lambda
\left\langle \mu \right. }-\left( 2r+5\right) \rho _{r}^{\lambda
\left\langle \mu \right. }\right] \sigma _{\lambda }^{\left. \nu
\right\rangle }  \notag \\
& +2\rho _{r}^{\lambda \left\langle \mu \right. }\omega _{\left.{\,}%
\right.\lambda }^{\left. \nu \right\rangle }-\Delta _{\alpha \beta }^{\mu
\nu }\nabla _{\lambda }\rho _{r-1}^{\alpha \beta \lambda }+(r-1)\rho
_{r-2}^{\mu \nu \lambda \kappa }\sigma _{\lambda \kappa }.
\label{tensor_rho}
\end{align}%
Here we introduced the generalized collision term%
\begin{align}
C_{r}^{\left\langle \mu _{1}\cdots \mu _{\ell }\right\rangle }& \equiv
\Delta _{\nu _{1}\cdots \nu _{\ell }}^{\mu _{1}\cdots \mu _{\ell
}}C_{r}^{\nu _{1}\ldots \nu _{\ell }}  \notag \\
& =\Delta _{\nu _{1}\cdots \nu _{\ell }}^{\mu _{1} \cdots \mu _{\ell }}\int
dKE_{\mathbf{k}}^{r}k^{\mu _{1}}\cdots k^{\mu _{\ell }}C\left[ f\right] .
\label{General_Coll_Int}
\end{align}%
All derivatives of $\alpha _{0}$ and $\beta _{0}$ that appear in the above
equations were replaced using the following equations, obtained from the
conservation laws~\eqref{bla1} and \eqref{bla2}, 
\begin{align}
\dot{\alpha}_{0}& =\frac{1}{D_{20}}\left[ -J_{30}\left( n_{0}\theta
+\partial _{\mu }V^{\mu }\right) +J_{20}\left( \varepsilon _{0}+P_{0}+\Pi
\right) \theta \right.  \notag \\
& \left. +J_{20}\left( \partial _{\mu }W^{\mu }-W^{\mu }\dot{u}_{\mu }-\pi
^{\mu \nu }\sigma _{\mu \nu }\right) \right] ,  \label{alpha_dot} \\
\dot{\beta}_{0}& =\frac{1}{D_{20}}\left[ -J_{20}\left( n_{0}\theta +\partial
_{\mu }V^{\mu }\right) +J_{10}\left( \varepsilon _{0}+P_{0}+\Pi \right)
\theta \right.  \notag \\
& \left. +J_{10}\left( \partial _{\mu }W^{\mu }-W^{\mu }\dot{u}_{\mu }-\pi
^{\mu \nu }\sigma _{\mu \nu }\right) \right] ,  \label{beta_dot} \\
\dot{u}^{\mu }& =\beta _{0}^{-1}\left( h_{0}^{-1}\nabla ^{\mu }\alpha
_{0}-\nabla ^{\mu }\beta _{0}\right) -\frac{h_{0}^{-1}}{n_{0}}\left( \Pi 
\dot{u}^{\mu }-\nabla ^{\mu }\Pi \right)  \notag \\
& -\frac{h_{0}^{-1}}{n_{0}}\left[ \frac{4}{3}W^{\mu }\theta +W_{\nu }\left(
\sigma ^{\mu \nu }-\omega ^{\mu \nu }\right) +\dot{W}^{\mu }+\Delta _{\nu
}^{\mu }\partial _{\lambda }\pi ^{\nu \lambda }\right] ,  \label{u_dot}
\end{align}%
where $h_{0}=(\varepsilon_{0}+P_{0})/n_{0}$. The coefficients $\alpha_{r}$
are functions of thermodynamic variables, 
\begin{align}
\alpha _{r}^{\left( 0\right) }& =\left( 1-r\right) I_{r1}-I_{r0} -\frac{n_{0}%
}{D_{20}}\left( h_{0} G_{2r} -G_{3r}\right) ,  \label{alpha_r_0} \\
\alpha _{r}^{\left( 1\right) }& =J_{r+1,1}-h_{0}^{-1}J_{r+2,1},
\label{alpha_r_1} \\
\alpha _{r}^{\left( 2\right) }& =I_{r+2,1}+\left( r-1\right) I_{r+2,2},
\label{alpha_r_2} \\
\alpha _{r}^{h}& = -\frac{\beta _{0}}{\varepsilon _{0}+P_{0}}J_{r+2,1},
\label{alpha_h}
\end{align}%
where we used the notation 
\begin{eqnarray}
I_{nq} &=&\frac{\left( -1\right) ^{q}}{\left( 2q+1\right) !!}\int dKE_{%
\mathbf{k}}^{n-2q}\left( \Delta ^{\alpha \beta }k_{\alpha }k_{\beta }\right)
^{q}f_{0\mathbf{k}},  \label{Inq} \\
J_{nq} &=&\frac{\left( -1\right) ^{q}}{\left( 2q+1\right) !!}\int dKE_{%
\mathbf{k}}^{n-2q}\left( \Delta ^{\alpha \beta }k_{\alpha }k_{\beta }\right)
^{q}f_{0\mathbf{k}}\tilde{f}_{0\mathbf{k}},\ \ \ \   \label{Jnq} \\
G_{nm} &=&J_{n0}J_{m0}-J_{n-1,0}J_{m+1,0},  \label{Gnm} \\
D_{nq} &=&J_{n+1,q}J_{n-1,q}-\left( J_{nq}\right) ^{2}.  \label{Dnq}
\end{eqnarray}

Thus, we have obtained an infinite set of coupled equations containing all
moments of the distribution function. Note that the derivation of these
equations is independent of the form of the expansion we introduced in the
previous subsection.

\section{The 14-moment approximation}

\label{14M}

In order to obtain the macroscopic equations of motion in terms of the
fluid-dynamical variables that appear in the particle four-current and
energy-momentum tensor, the generic hierarchy of the coupled moment
equations must be truncated. To this end, Israel and Stewart made the
so-called 14-moment approximation: they truncated the expansion of the
distribution function and matched the non-equilibrium corrections to the
dissipative currents $\Pi $, $V^{\mu }$, $W^{\mu }$, and $\pi ^{\mu \nu }$.

In this section, we show how the 14-moment approximation emerges from the
general moment expansion presented in Sec.\ \ref{General_eqs}. First, we
neglect irreducible tensor moments of rank higher than two, i.e., $\rho
_{r}^{\mu _{1}\cdots \mu _{\ell }}=0$ for $\ell \geq 3$ in Eqs.\ (\ref%
{vector_rho},\ref{tensor_rho}). Such irreducible moments cannot be
constructed purely from first-order gradients of equilibrium fields~\cite%
{Denicol:2012cn}. This means that they lead to terms that are of higher
order in gradients or contain higher powers of dissipative quantities in the
equations of motion.

Next, in the expansion (\ref{main_f_expansion}) of the distribution function
we include the first three scalar moments, namely $\rho _{0}=-3\Pi /m^{2}$, $%
\rho _{1}=0$, and $\rho _{2}=0$ (the last two scalar moments vanish due to
the matching condition, but must be included since they were used to define $%
\alpha _{0}$ and $\beta _{0}$), the first two vector moments, $\rho
_{0}^{\mu }=V^{\mu }$ and $\rho _{1}^{\mu }=W^{\mu },$ and the first
second-rank tensor moment $\rho _{0}^{\mu \nu }=\pi ^{\mu \nu }$. This
implies that $N_{0}=2$, $N_{1}=1$, and $N_{2}=0$, while all other moments
appearing in the expansion are dropped. Choosing either the Eckart or the
Landau frame, we can eliminate either $V^{\mu }$ or $W^{\mu }$,
respectively. Let us note that, so far, this approach is completely
equivalent to the matching procedure of Israel and Stewart. The ambiguity in
the transport coefficients emerges only at a later stage, when choosing
moments of the Boltzmann equation to supply the equations of motion for the
dissipative currents.

The restriction to the aforementioned moments\ affects that $\lambda _{%
\mathbf{k}}$, $\lambda _{\mathbf{k}}^{\left\langle \mu \right\rangle }$, and 
$\lambda _{\mathbf{k}}^{\left\langle \mu \nu \right\rangle }$ from Eq.\ (\ref%
{expansion2}) can be expressed solely in terms of $\Pi $, $V^{\mu }$, $%
W^{\mu }$, and $\pi ^{\mu \nu }$, 
\begin{align}
\lambda _{\mathbf{k}}& \equiv \sum_{n=0}^{N_{0}}c_{n}P_{\mathbf{k}n}^{\left(
0 \right) }\simeq c_{0}P_{\mathbf{k}0}^{\left( 0\right) }+c_{1}P_{\mathbf{k}%
1}^{\left( 0\right) }+c_{2}P_{\mathbf{k}2}^{\left( 0\right) }, \\
\lambda _{\mathbf{k}}^{\left\langle \mu \right\rangle }& \equiv
\sum_{n=0}^{N_{1}}c_{n}^{\left\langle \mu \right\rangle }P_{\mathbf{k}%
n}^{\left( 1 \right) }\simeq c_{0}^{\left\langle \mu \right\rangle }P_{%
\mathbf{k}0}^{\left( 1\right) }+c_{1}^{\left\langle \mu \right\rangle }P_{%
\mathbf{k}1}^{\left( 1\right) }, \\
\lambda _{\mathbf{k}}^{\left\langle \mu \nu \right\rangle }& \equiv
\sum_{n=0}^{N_{2}}c_{n}^{\left\langle \mu \nu \right\rangle }P_{\mathbf{k}%
n}^{\left( 2 \right) }\simeq c_{0}^{\left\langle \mu \nu \right\rangle }P_{%
\mathbf{k}0}^{\left( 2\right) },
\end{align}%
where the tensors $c_{n}^{\left\langle \mu _{1}\cdots \mu _{\ell
}\right\rangle }$ are given by Eq. (\ref{c_coeff}), while those which do not
appear in the above equations are set to zero. According to Eq. (\ref%
{c_coeff}) the scalars $c_{0}$, $c_{1}$, and $c_{2}$ are proportional to the
bulk viscous pressure, 
\begin{align}
c_{0}& =-\frac{3\Pi }{m^{2}}a_{00}^{(0)}\mathcal{N}^{\left( 0\right) },\  \\
c_{1}& =-\frac{3\Pi }{m^{2}}a_{10}^{(0)}\mathcal{N}^{\left( 0\right) },\  \\
c_{2}& =-\frac{3\Pi }{m^{2}}a_{20}^{(0)}\mathcal{N}^{\left( 0\right) }.
\end{align}%
The vectors $c_{0}^{\left\langle \mu \right\rangle }$ and $%
c_{1}^{\left\langle \mu \right\rangle }$ are given by a linear combination
of particle and energy-momentum diffusion currents,%
\begin{align}
c_{0}^{\left\langle \mu \right\rangle }& =V^{\mu }a_{00}^{(1)}\mathcal{N}%
^{\left( 1\right) },\  \\
c_{1}^{\left\langle \mu \right\rangle }& =V^{\mu }a_{10}^{(1)}\mathcal{N}%
^{\left( 1\right) }+W^{\mu }a_{11}^{(1)}\mathcal{N}^{\left( 1\right) },
\end{align}%
while $c_{0}^{\left\langle \mu \nu \right\rangle }$ is proportional to the
shear-stress tensor, 
\begin{equation}
c_{0}^{\left\langle \mu \nu \right\rangle }=\pi ^{\mu \nu }a_{00}^{(2)}\frac{%
\mathcal{N}^{\left( 2\right) }}{2}.
\end{equation}

Let us recall Eq.\ (\ref{Poly}) and for any$\ \ell \geq 0$ we set 
\begin{equation}
P_{\mathbf{k}0}^{\left( \ell \right) }\equiv a_{00}^{(\ell )}=1,
\end{equation}%
while 
\begin{align}
P_{\mathbf{k}1}^{\left( 0\right) }&= a_{11}^{\left( 0\right) }E_{\mathbf{k}%
}+a_{10}^{\left( 0\right) }, \\
P_{\mathbf{k}1}^{\left( 1\right) }& = a_{11}^{\left( 1\right) }E_{\mathbf{k}%
}+a_{10}^{\left( 1\right) }, \\
P_{\mathbf{k}2}^{\left( 0\right) }& = a_{22}^{\left( 0\right) }E_{\mathbf{k}%
}^{2}+a_{21}^{\left( 0\right) }E_{\mathbf{k}}+a_{20}^{\left( 0\right) }.
\end{align}%
The orthonormality condition (\ref{conditions}) implies that the
normalization constant is 
\begin{equation}
\mathcal{N}^{\left( \ell \right) }=\left( J_{2\ell ,\ell }\right) ^{-1},
\end{equation}%
and 
\begin{align}
\frac{a_{10}^{(0)}}{a_{11}^{(0)}}& =-\frac{J_{10}}{J_{00}},\ \left(
a_{11}^{(0)}\right) ^{2}=\frac{J_{00}^{2}}{D_{10}}, \\
\frac{a_{21}^{\left( 0\right) }}{a_{22}^{\left( 0\right) }}& =\frac{G_{12}}{%
D_{10}},\ \frac{a_{20}^{\left( 0\right) }}{a_{22}^{\left( 0\right) }}=\frac{%
D_{20}}{D_{10}}, \\
\left( a_{22}^{\left( 0\right) }\right) ^{2}& =\frac{J_{00}D_{10}}{%
J_{20}D_{20}+J_{30}G_{12}+J_{40}D_{10}}, \\
\frac{a_{10}^{\left( 1\right) }}{a_{11}^{\left( 1\right) }}& =-\frac{J_{31}}{%
J_{21}},\ \left( a_{11}^{\left( 1\right) }\right) ^{2}=\frac{J_{21}^{2}}{%
D_{31}}.
\end{align}%
Furthermore, using the orthogonality relation (\ref{orthogonality1})
together with Eqs.\ (\ref{expansion1}-\ref{Poly}) one can easily show that 
\begin{equation}
\rho _{r}^{\mu _{1}\cdots \mu _{\ell }}=\ell !\sum_{n=0}^{N_{\ell
}}\sum_{m=0}^{n}c_{n}^{\left\langle \mu _{1}\cdots \mu _{\ell }\right\rangle
}a_{nm}^{(\ell )}\ J_{r+m+2\ell ,\ell }.
\end{equation}%
Applying the truncation scheme required by the 14-moment approximation we
obtain that all scalar moments, $\rho _{r}$, become proportional to the bulk
viscous pressure $\Pi $, 
\begin{equation}
\rho _{r}\equiv \sum_{n=0}^{N_{0}}\sum_{m=0}^{n}c_{n}a_{nm}^{(0)}\
J_{r+m,0}=\gamma _{r}^{\Pi }\Pi .  \label{rho_scalar_final}
\end{equation}%
Similarly, all vector moments, $\rho _{r}^{\mu }$, are proportional to a
linear combination of $V^{\mu }$ and $W^{\mu }$, 
\begin{equation}
\rho _{r}^{\mu }\equiv \sum_{n=0}^{N_{1}}\sum_{m=0}^{n}c_{n}^{\left\langle
\mu \right\rangle }a_{nm}^{(1)}\ J_{r+m+2,1}=\gamma _{r}^{V}V^{\mu }+\gamma
_{r}^{W}W^{\mu },  \label{rho_vector_final}
\end{equation}%
and, finally, $\rho _{r}^{\mu \nu }$ is proportional to $\pi ^{\mu \nu }$,%
\begin{equation}
\rho _{r}^{\mu \nu }\equiv
\sum_{n=0}^{N_{2}}\sum_{m=0}^{n}c_{n}^{\left\langle \mu \nu \right\rangle
}a_{nm}^{(2)}\ J_{r+m+4,2}=\gamma _{r}^{\pi }\pi ^{\mu \nu }.
\label{rho_tensor_final}
\end{equation}

Now, using the previously obtained results we prove that 
\begin{eqnarray}
\gamma _{r}^{\Pi } &=&\mathcal{A}_{\Pi }J_{r0}+\mathcal{B}_{\Pi }J_{r+1,0}+%
\mathcal{C}_{\Pi }J_{r+2,0},  \label{gamma_Pi} \\
\gamma _{r}^{V} &=&\mathcal{A}_{V}J_{r+2,1}+\mathcal{B}_{V}J_{r+3,1},
\label{gamma_V} \\
\gamma _{r}^{W} &=&\mathcal{A}_{W}J_{r+2,1}+\mathcal{B}_{W}J_{r+3,1},
\label{gamma_W} \\
\gamma _{r}^{\pi } &=&2\mathcal{A}_{\pi }J_{r+4,2},  \label{gamma_pi}
\end{eqnarray}%
where 
\begin{align}
\mathcal{A}_{\Pi }& =-\frac{3}{m^{2}}\frac{D_{30}}{
J_{20}D_{20}+J_{30}G_{12}+J_{40}D_{10} },  \label{A_Pi} \\
\mathcal{B}_{\Pi }& =-\frac{3}{m^{2}}\frac{G_{23}}{
J_{20}D_{20}+J_{30}G_{12}+J_{40}D_{10}}, \\
\mathcal{C}_{\Pi }& =-\frac{3}{m^{2}}\frac{D_{20}}{
J_{20}D_{20}+J_{30}G_{12}+J_{40}D_{10}},  \label{C_Pi} \\
\mathcal{A}_{V}& =\frac{J_{41}}{D_{31}},\ \mathcal{A}_{W}=-\frac{J_{31}}{%
D_{31}}, \\
\mathcal{B}_{V}& =-\frac{J_{31}}{D_{31}},\ \mathcal{B}_{W}=\frac{J_{21}}{%
D_{31}},  \label{B_V_W} \\
\mathcal{A}_{\pi }& =\frac{1}{2J_{42}}.  \label{A_pi}
\end{align}%
We remark that, since the matching conditions were already imposed, one can
prove that $\gamma _{1}^{\Pi }=\gamma _{2}^{\Pi }=0$.

Equations (\ref{rho_scalar_final}), (\ref{rho_vector_final}), and (\ref%
{rho_tensor_final}) are the main result of the 14-moment approximation. Such
relations guarantee that any irreducible moment of the distribution function
can be expressed in terms of the dissipative currents appearing in $N^{\mu}$
and $T^{\mu \nu}$. This is also what Israel and Stewart achieved by their
matching procedure. Consequently, a closed set of fluid-dynamical equations
can always be derived. This happens because the reduction of dynamical
variables was done directly in the single-particle distribution. On the
other hand, this truncation also leads to an ambiguity in the derivation of
fluid-dynamical equations since, for example, the equation of motion for the
bulk viscous pressure can be obtained from $\rho _{r}$ for any $r$. We will
come back to this point in Sec.\ \ref{choice_of_moment}.

\subsection{The collision term}

\label{14M_coll}

In order to express the collision term (\ref{General_Coll_Int}) in terms of
the fundamental fluid variables, $C[f]$ is linearized in deviations from the
equilibrium distribution function. Substituting Eq.\ (\ref{small_correc})
into the linearized collision term, we obtain 
\begin{align}
C_{r-1}^{\mu _{1}\cdots \mu _{\ell }}& =\frac{1}{\nu }\int dKdK^{\prime
}dPdP^{\prime }W_{\mathbf{kk}\prime \rightarrow \mathbf{pp}\prime }f_{0%
\mathbf{k}}f_{0\mathbf{k}\prime }\tilde{f}_{0\mathbf{p}}\tilde{f}_{0\mathbf{p%
}\prime }  \notag \\
& \times E_{\mathbf{k}}^{r-1}k^{\mu _{1}}\cdots k^{\mu _{\ell }}\left( \phi
_{\mathbf{p}}+\phi _{\mathbf{p}^{\prime }}-\phi _{\mathbf{k}}-\phi _{\mathbf{%
k}^{\prime }}\right) ,  \label{linear_collision}
\end{align}%
where the $\phi $'s are given in Eq.\ (\ref{expansion1}). In order to
specify $C_{r-1}$, $C_{r-1}^{\left\langle \alpha \right\rangle }$, and $%
C_{r-1}^{\left\langle \alpha \beta \right\rangle }$ for the general
equations of motion (\ref{scalar_rho}, \ref{vector_rho}, \ref{tensor_rho})
we project the collision term as 
\begin{align}
C_{r-1}& =u_{\mu _{1}}\cdots u_{\mu _{\ell }}C_{r-\ell -1}^{\mu _{1}\cdots
\mu _{\ell }},  \label{C_scalar} \\
C_{r-1}^{\left\langle \alpha \right\rangle }& =\Delta _{\mu _{1}}^{\alpha
}u_{\mu _{2}}\cdots u_{\mu _{\ell }}C_{r-\ell }^{\mu _{1}\cdots \mu _{\ell
}},  \label{C_vector} \\
C_{r-1}^{\left\langle \alpha \beta \right\rangle }& =\Delta _{\mu _{1}\mu
_{2}}^{\alpha \beta }u_{\mu _{3}}\cdots u_{\mu _{\ell }}C_{r-\ell +1}^{\mu
_{1}\cdots \mu _{\ell }}.  \label{C_tensor}
\end{align}

In the 14-moment approximation we start by substituting Eqs.\ (\ref%
{expansion1}) and (\ref{expansion2}) into Eq.\ (\ref{linear_collision}) and
obtain 
\begin{eqnarray}
C_{r-1} &=&\mathcal{C}_{\Pi \chi }\Pi X_{r-3,1}, \\
C_{r-1}^{\left\langle \mu \right\rangle } &=&\mathcal{B}_{V}V^{\mu
}X_{r-2,3}+\mathcal{B}_{W}W^{\mu }X_{r-2,3}, \\
C_{r-1}^{\left\langle \mu \nu \right\rangle } &=&\mathcal{A}_{\pi }\pi ^{\mu
\nu }X_{r-1,4}.
\end{eqnarray}%
Here $X_{r,1}$, $X_{r,3}$ and $X_{r,4}$ are coefficients of the following
rank-$4$ collision tensor,%
\begin{align}
X_{r}^{\mu \nu \alpha \beta }& =\frac{1}{\nu }\int dKdK^{\prime
}dPdP^{\prime }W_{\mathbf{kk}\prime \rightarrow \mathbf{pp}\prime }f_{0%
\mathbf{k}}f_{0\mathbf{k}\prime }\tilde{f}_{0\mathbf{p}}\tilde{f}_{0\mathbf{p%
}\prime }  \notag \\
& \times E_{\mathbf{k}}^{r}k^{\mu }k^{\nu }\left( p^{\alpha }p^{\beta
}+p^{\prime \alpha }p^{\prime \beta }-k^{\alpha }k^{\beta }-k^{\prime \alpha
}k^{\prime \beta }\right) .  \label{X_4rank_tens_def}
\end{align}%
This collision tensor is symmetric upon the interchange of indices $\left(
\mu ,\nu \right) $ and $\left( \alpha ,\beta \right) $, and also traceless
for the latter indices, 
\begin{equation}
X_{r}^{\mu \nu \alpha \beta }=X_{r}^{\left( \mu \nu \right) \left( \alpha
\beta \right) },\ X_{r}^{\mu \nu \alpha \beta }g_{\alpha \beta }=0.
\end{equation}%
These properties lead to a spatially isotropic tensor constructed using the
four-velocity $u^{\mu }$, the transverse projection $\Delta ^{\mu \nu }$,
and different scalar coefficients $X_{r,i}$ as 
\begin{align}
X_{r}^{\mu \nu \alpha \beta }& =\left( X_{r,1}u^{\mu }u^{\nu }+X_{r,2}\Delta
^{\mu \nu }\right) \left( u^{\alpha }u^{\beta }-\frac{1}{3}\Delta ^{\alpha
\beta }\right)  \notag  \label{X_4rank_tens} \\
& +4X_{r,3}u^{\left( \mu \right. }\Delta ^{\left. \nu \right) \left( \alpha
\right. }u^{\left. \beta \right) }+X_{r,4}\Delta ^{\mu \nu \alpha \beta },
\end{align}%
where the coefficients of the above decomposition are obtained from the
following contractions, 
\begin{align}
X_{r,1}& =X_{r}^{\mu \nu \alpha \beta }u_{\mu }u_{\nu }u_{\alpha }u_{\beta
}\ ,  \label{X1_tens} \\
X_{r,2}& =\frac{1}{3}X_{r}^{\mu \nu \alpha \beta }\Delta _{\mu \nu
}u_{\alpha }u_{\beta }\ ,  \label{X2_tens} \\
X_{r,3}& =\frac{1}{3}X_{r}^{\mu \nu \alpha \beta }u_{\mu }\Delta _{\nu
\alpha }u_{\beta }\ ,  \label{X3_tens} \\
X_{r,4}& =\frac{1}{5}X_{r}^{\mu \nu \alpha \beta }\Delta _{\mu \nu \alpha
\beta }\ .  \label{X4_tens}
\end{align}

The evaluation of the coefficients $X_{r,i}$ requires the detailed knowledge
of the differential cross section. As an example, these functions are
evaluated for a classical gas of massless particles with constant cross
section in Appendix \ref{relax_times}.

\subsection{Equations of motion}

\label{Relax_equations}

In order to close the conservation laws Eqs.\ (\ref{bla1}) and (\ref{bla2})
we need additional equations for the dissipative currents which can be
obtained from the exact equations of motion, Eqs.\ (\ref{scalar_rho}, \ref%
{vector_rho}, \ref{tensor_rho}).

For any $r\geq 0$, Eq.\ (\ref{scalar_rho}) for the scalar moment leads to an
equation of motion for the bulk viscous pressure. Replacing$\ \rho
_{r}=\gamma _{r}^{\Pi }\Pi $ according to Eq.\ (\ref{rho_scalar_final}) in
Eq.\ (\ref{scalar_rho}) and collecting all terms we obtain the so-called
relaxation equation of the bulk viscous pressure,%
\begin{align}
\dot{\Pi}& =-\frac{\Pi }{\tau _{\Pi }^{r}}-\frac{\zeta ^{r}}{\tau _{\Pi }^{r}%
}\theta +\tau _{\Pi W}^{r}W_{\mu }\dot{u}^{\mu }+\tau _{\Pi V}^{r}V_{\mu }%
\dot{u}^{\mu }  \notag \\
& -\ell _{\Pi W}^{r}\partial _{\mu }W^{\mu }-\ell _{\Pi V}^{r}\partial _{\mu
}V^{\mu }-\delta _{\Pi \Pi }^{r}\Pi \theta   \notag \\
& +\lambda _{\Pi W}^{r}W^{\mu }\nabla _{\mu }\alpha _{0}+\lambda _{\Pi
V}^{r}V^{\mu }\nabla _{\mu }\alpha _{0}+\lambda _{\Pi \pi }^{r}\pi ^{\mu \nu
}\sigma _{\mu \nu }.  \label{relax_bulk}
\end{align}%
Here, we introduced the relaxation time of the bulk viscous pressure, $\tau
_{\Pi }^{r}$, and the bulk viscosity coefficient $\zeta ^{r}$ as%
\begin{equation}
\tau _{\Pi }^{r}=-\frac{\gamma _{r}^{\Pi }\mathcal{C}_{\Pi }^{-1}}{X_{r-3,1}}%
,\;\;\;\;\zeta ^{r}=-\tau _{\Pi }^{r}\frac{\alpha _{r}^{\left( 0\right) }}{%
\gamma _{r}^{\Pi }},
\end{equation}%
where $\mathcal{C}_{\Pi }$ was defined in Eq.\ (\ref{C_Pi}). Similarly, $%
\gamma _{r}^{\Pi }$ was defined in Eq.\ (\ref{gamma_Pi}), $X_{r,1}$ in Eq.\ (%
\ref{X1_tens}), while $\alpha _{r}^{\left( 0\right) }$ was given in Eq.\ (%
\ref{alpha_r_0}). The other transport coefficients in Eq.\ (\ref{relax_bulk}%
) are defined as 
\begin{align}
\tau _{\Pi W}^{r}& =\frac{1}{\gamma _{r}^{\Pi }}\left[ \left( r-1\right)
\gamma _{r-1}^{W}+\beta _{0}\frac{\partial \gamma _{r-1}^{W}}{\partial \beta
_{0}}+\frac{G_{2r}}{D_{20}}\right] , \\
\tau _{\Pi V}^{r}& =\frac{1}{\gamma _{r}^{\Pi }}\left[ \left( r-1\right)
\gamma _{r-1}^{V}+\beta _{0}\frac{\partial \gamma _{r-1}^{V}}{\partial \beta
_{0}}\right] , \\
\lambda _{\Pi W}^{r}& =-\frac{1}{\gamma _{r}^{\Pi }}\left( \frac{\partial
\gamma _{r-1}^{W}}{\partial \alpha _{0}}+h_{0}^{-1}\frac{\partial \gamma
_{r-1}^{W}}{\partial \beta _{0}}\right) , \\
\lambda _{\Pi V}^{r}& =-\frac{1}{\gamma _{r}^{\Pi }}\left( \frac{\partial
\gamma _{r-1}^{V}}{\partial \alpha _{0}}+h_{0}^{-1}\frac{\partial \gamma
_{r-1}^{V}}{\partial \beta _{0}}\right) , \\
\ell _{\Pi W}^{r}& =\frac{1}{\gamma _{r}^{\Pi }}\left( \gamma _{r-1}^{W}+%
\frac{G_{2r}}{D_{20}}\right) , \\
\ell _{\Pi V}^{r}& =\frac{1}{\gamma _{r}^{\Pi }}\left( \gamma _{r-1}^{V}-%
\frac{G_{3r}}{D_{20}}\right) , \\
\lambda _{\Pi \pi }^{r}& =\frac{1}{\gamma _{r}^{\Pi }}\left[ \left(
r-1\right) \gamma _{r-2}^{\pi }+\frac{G_{2r}}{D_{20}}\right] , \\
\delta _{\Pi \Pi }^{r}& =\frac{n_{0}D_{20}^{-1}}{\gamma _{r}^{\Pi }}\left[
\left( J_{20}\frac{\partial \gamma _{r}^{\Pi }}{\partial \alpha _{0}}+J_{10}%
\frac{\partial \gamma _{r}^{\Pi }}{\partial \beta _{0}}\right) h_{0}\right. 
\notag \\
& \left. -J_{30}\frac{\partial \gamma _{r}^{\Pi }}{\partial \alpha _{0}}%
-J_{20}\frac{\partial \gamma _{r}^{\Pi }}{\partial \beta _{0}}\right] -\frac{%
1}{3\gamma _{r}^{\Pi }}\left. \bigg[\left( r-1\right) m^{2}\gamma
_{r-2}^{\Pi }\right.   \notag \\
& \left. -\left( r+2\right) \gamma _{r}^{\Pi }-3\frac{G_{2r}}{D_{20}}\bigg]%
\right. .
\end{align}%
Note that these coefficients are independent of the collision integral. We
also point out that here we follow the notation of Refs.\ \cite%
{Betz:2008me,Betz:2010cx} for the coefficients. Furthermore the choice of
the LRF eliminates terms involving either $V^{\mu }$ (for Eckart's choice)
or $W^{\mu }$ (for Landau's choice).

In the very same manner we get a relaxation equation for both the particle
diffusion current and the energy-momentum diffusion current. This equation
follows from Eq. (\ref{vector_rho}) using $\rho _{r}^{\mu }=\gamma
_{r}^{V}V^{\mu }+\gamma _{r}^{W}W^{\mu }$, where rank-$3$ tensors $\rho
_{r}^{\mu \nu \lambda }$ for any $r\geq 0$ were neglected. Thus, after some
calculations we obtain 
\begin{align}
\dot{V}^{\mu }+\psi _{r}^{W}\dot{W}^{\mu }& =-\frac{V^{\mu }}{\tau _{V}^{r}}%
-\psi _{r}^{W}\frac{W^{\mu }}{\tau _{W}^{r}}+\frac{\kappa _{q}^{r}}{\tau
_{V}^{r}\beta _{0}^{2}h_{0}^{2}}\nabla ^{\mu }\alpha _{0}  \notag \\
& +\psi _{r}^{W}W_{\nu }\omega ^{\mu \nu }+V_{\nu }\omega ^{\mu \nu }  \notag
\\
& -\psi _{r}^{W}\lambda _{WW}^{r}W_{\nu }\sigma ^{\mu \nu }+\lambda
_{VV}^{r}V_{\nu }\sigma ^{\mu \nu }  \notag \\
& -\psi _{r}^{W}\delta _{WW}^{r}W^{\mu }\theta +\delta _{VV}^{r}V^{\mu
}\theta   \notag \\
& -\tau _{q\Pi }^{r}\Pi \dot{u}^{\mu }-\tau _{q\pi }^{r}\pi ^{\mu \nu }\dot{u%
}_{\nu }  \notag \\
& +\ell _{q\Pi }^{r}\nabla ^{\mu }\Pi -\ell _{q\pi }^{r}\Delta _{\alpha
}^{\mu }\partial _{\nu }\pi ^{\alpha \nu }  \notag \\
& +\lambda _{q\Pi }^{r}\Pi \nabla ^{\mu }\alpha _{0}+\lambda _{q\pi }^{r}\pi
^{\mu \nu }\nabla _{\nu }\alpha _{0},  \label{relax_heat}
\end{align}%
where we defined the relaxation time of the particle diffusion current, $%
\tau _{V}^{r}$, of the energy-momentum diffusion current, $\tau _{W}^{r}$,
and the heat conductivity coefficient $\kappa _{q}^{r}$, 
\begin{align}
\tau _{V}^{r}& =-\frac{\gamma _{r}^{V}}{\mathcal{B}_{V}X_{r-2,3}}, \\
\tau _{W}^{r}& =-\frac{\gamma _{r}^{W}+\alpha _{r}^{h}}{\mathcal{B}%
_{W}X_{r-2,3}},\   \label{tau_q} \\
\kappa _{q}^{r}& =\tau _{V}^{r}\frac{\alpha _{r}^{\left( 1\right) }}{\gamma
_{r}^{V}}h_{0}^{2}\beta _{0}^{2}.
\end{align}%
Furthermore, 
\begin{equation}
\psi _{r}^{W}=\frac{\gamma _{r}^{W}+\alpha _{r}^{h}}{\gamma _{r}^{V}},
\end{equation}%
and $\mathcal{B}_{V}$, $\mathcal{B}_{W}$ were defined in Eqs.\ (\ref{B_V_W}%
), such that $\mathcal{B}_{V}=-h_{0}\mathcal{B}_{W}$, while $\gamma _{r}^{V}$%
, $\gamma _{r}^{W}$ were defined in Eqs.\ (\ref{gamma_V},\ref{gamma_W}). In
addition, $X_{r,3}$ was defined in Eq.\ (\ref{X3_tens}), while $\alpha
_{r}^{\left( 1\right) }$ and $\alpha _{r}^{h}$ were given in Eqs.\ (\ref%
{alpha_r_1}) and (\ref{alpha_h}). The two relaxation times are not
independent but are related to each other as $\tau _{W}^{r}=-h_{0}\psi
_{r}^{W}\tau _{V}^{r}$. The relaxation equation is written such that it is
straightforward to rewrite it in either the Eckart or the Landau frame.
Moreover, using the definition (\ref{heatflow}) of the heat flow it is clear
that the above equation is a relaxation equation for this quantity.

The coefficients in Eq.\ (\ref{relax_heat}) which only exist in either the
Eckart or Landau frame are 
\begin{align}
\lambda _{WW}^{r}& =-\frac{1}{5\gamma _{r}^{V}\psi _{r}^{W}}\left[
m^{2}\left( 2r-2\right) \gamma _{r-2}^{W}\right.  \notag \\
& \left. -\left( 2r+3\right) \gamma _{r}^{W}-5\alpha _{r}^{h}\right] , \\
\delta _{WW}^{r}& =\frac{n_{0}D_{20}^{-1}}{\gamma _{r}^{V}\psi _{r}^{W}}%
\left[ \left( J_{20}\frac{\partial \gamma _{r}^{W}}{\partial \alpha _{0}}%
+J_{10}\frac{\partial \gamma _{r}^{W}}{\partial \beta _{0}}\right)
h_{0}\right.  \notag \\
& \left. -J_{30}\frac{\partial \gamma _{r}^{W}}{\partial \alpha _{0}}-J_{20}%
\frac{\partial \gamma _{r}^{W}}{\partial \beta _{0}}\right] -\frac{1}{%
3\gamma _{r}^{V}\psi _{r}^{W}}\left[ m^{2}\left( r-1\right) \gamma
_{r-2}^{W}\right.  \notag \\
& \left. -\left( r+3\right) \gamma _{r}^{W}-4\alpha _{r}^{h}\right] ,
\label{delta_q}
\end{align}%
and 
\begin{align}
\lambda _{VV}^{r}& =\frac{1}{5\gamma _{r}^{V}}\left[ m^{2}\left( 2r-2\right)
\gamma _{r-2}^{V}-\left( 2r+3\right) \gamma _{r}^{V}\right] , \\
\delta _{VV}^{r}& =- \frac{n_{0}D_{20}^{-1}}{\gamma _{r}^{V}}\left[ \left(
J_{20}\frac{\partial \gamma _{r}^{V}}{\partial \alpha _{0}}+J_{10}\frac{%
\partial \gamma _{r}^{V}}{\partial \beta _{0}}\right) h_{0}\right.  \notag \\
& \left. -J_{30}\frac{\partial \gamma _{r}^{V}}{\partial \alpha _{0}}-J_{20}%
\frac{\partial \gamma _{r}^{V}}{\partial \beta _{0}}\right]  \notag \\
& +\frac{1}{3\gamma _{r}^{V}}\left[ m^{2}\left( r-1\right) \gamma
_{r-2}^{V}V^{\mu }-\left( r+3\right) \gamma _{r}^{V}\right] ,
\end{align}%
while terms and coefficients which are not affected by either choice of the
LRF are 
\begin{align}
\tau _{q\pi }^{r}& =-\frac{1}{\gamma _{r}^{V}}\left[ \left( r-1\right)
\gamma _{r-1}^{\pi }+\beta _{0}\frac{\partial \gamma _{r-1}^{\pi }}{\partial
\beta _{0}}\right] , \\
\tau _{q\Pi }^{r}& =-\frac{1}{3\gamma _{r}^{V}}\left[ r m^{2}\gamma
_{r-1}^{\Pi }-\left( r+3\right) \gamma _{r+1}^{\Pi }\right.  \notag \\
& \left. -3\alpha _{r}^{h}+m^{2}\beta _{0}\frac{\partial \gamma _{r-1}^{\Pi }%
}{\partial \beta _{0}}-\beta _{0}\frac{\partial \gamma _{r+1}^{\Pi }}{%
\partial \beta _{0}}\right] , \\
\ell_{q\pi }^{r}& =\frac{1}{\gamma _{r}^{V}}\left( \alpha _{r}^{h}+\gamma
_{r-1}^{\pi }\right) , \\
\ell_{q\Pi }^{r}& =\frac{1}{\gamma _{r}^{V}}\left[ \alpha _{r}^{h}-\frac{1}{3%
}\left( m^{2}\gamma _{r-1}^{\Pi }-\gamma _{r+1}^{\Pi }\right) \right] , \\
\lambda _{q\pi }^{r}& =-\frac{1}{\gamma _{r}^{V}}\left( \frac{\partial
\gamma _{r-1}^{\pi }}{\partial \alpha _{0}}+h_{0}^{-1}\frac{\partial \gamma
_{r-1}^{\pi }}{\partial \beta _{0}}\right) ,  \label{lambda_pi_q} \\
\lambda _{q\Pi }^{r}& =-\frac{1}{3\gamma _{r}^{V}}\left[ m^{2}\left( \frac{%
\partial \gamma _{r-1}^{\Pi }}{\partial \alpha _{0}}+h_{0}^{-1}\frac{%
\partial \gamma _{r-1}^{\Pi }}{\partial \beta _{0}}\right) \right.  \notag \\
& \left. -\left( \frac{\partial \gamma _{r+1}^{\Pi }}{\partial \alpha _{0}}%
+h_{0}^{-1}\frac{\partial \gamma _{r+1}^{\Pi }}{\partial \beta _{0}}\right) %
\right] .
\end{align}

The relaxation equation of the shear-stress tensor follows from Eq.\ (\ref%
{tensor_rho}) by replacing $\rho _{r}^{\left\langle \mu \nu \right\rangle
}=\gamma _{r}^{\pi }\pi ^{\mu \nu }$, and neglecting rank-3 and rank-4
tensors, $\rho ^{\mu \nu \lambda }=0$ and $\rho ^{\mu \nu \lambda \kappa }=0$%
, 
\begin{align}
\dot{\pi}^{\left\langle \mu \nu \right\rangle }& =-\frac{\pi ^{\mu \nu }}{%
\tau _{\pi }^{r}}+\frac{2\eta ^{r}}{\tau _{\pi }^{r}}\sigma ^{\mu \nu
}+2\lambda _{\pi \Pi }^{r}\Pi \sigma ^{\mu \nu }  \notag \\
& +2\tau _{\pi W}^{r}W^{\left\langle \mu \right. }\dot{u}^{\left. \nu
\right\rangle }+2\tau _{\pi V}^{r}V^{\left\langle \mu \right. }\dot{u}%
^{\left. \nu \right\rangle }  \notag \\
& +2 \ell_{\pi W}^{r}\nabla ^{\left\langle \mu \right. }W^{\left. \nu
\right\rangle }+2 \ell_{\pi V}^{r}\nabla ^{\left\langle \mu \right.
}V^{\left. \nu \right\rangle }  \notag \\
& -2\lambda _{\pi W}^{r}W^{\left\langle \mu \right. }\nabla ^{\left. \nu
\right\rangle }\alpha _{0}-2\lambda _{\pi V}^{r}V^{\left\langle \mu \right.
}\nabla ^{\left. \nu \right\rangle }\alpha _{0}  \notag \\
& -2\lambda _{\pi \pi }^{r}\pi _{\alpha }^{\left\langle \mu \right. }\sigma
^{\left. \nu \right\rangle \alpha }+2\pi _{\alpha }^{\left\langle \mu
\right. }\omega ^{\left. \nu \right\rangle \alpha }-2\delta _{\pi \pi
}^{r}\pi ^{\mu \nu }\theta ,  \label{relax_stress}
\end{align}%
where we defined the relaxation time $\tau _{\pi }^{r}$ for the shear-stress
tensor and the shear viscosity coefficient $\eta ^{r}$, 
\begin{equation}  \label{tau_pi}
\tau _{\pi }^{r} =- \frac{\gamma _{r}^{\pi }}{\mathcal{A}_{\pi }X_{r-1,4}},
\;\;\;\; \eta ^{r} =\tau _{\pi }^{r}\frac{\alpha _{r}^{\left( 2\right) }}{%
\gamma _{r}^{\pi }}.
\end{equation}%
Here, $\mathcal{A}_{\pi }$ was given in Eq.\ (\ref{A_pi}), $\gamma _{r}^{\pi
}$ in Eq.\ (\ref{gamma_pi}), $X_{r,4}$ in Eq.\ (\ref{X4_tens}), while $%
\alpha _{r}^{\left( 2\right) }$ was quoted in Eq.\ (\ref{alpha_r_2}). The
other coefficients in Eq.\ (\ref{relax_stress}) are 
\begin{align}
\lambda _{\pi \Pi }^{r}& =\frac{1}{15\gamma _{r}^{\pi }}\left[ \left(
r-1\right) m^{4}\gamma _{r-2}^{\Pi }-\left( 2r+3\right) m^{2}\gamma
_{r}^{\Pi }\right.  \notag \\
& \left. +(r+4)\gamma _{r+2}^{\Pi }\right] , \\
\tau _{\pi W}^{r}& =\frac{1}{5\gamma _{r}^{\pi }}\left[ r m^{2}\gamma
_{r-1}^{W}+m^{2}\beta _{0}\frac{\partial \gamma _{r-1}^{W}}{\partial \beta
_{0}}\right.  \notag \\
& \left. -\left( r+5\right) \gamma _{r+1}^{W}-\beta _{0}\frac{\partial
\gamma _{r+1}^{W}}{\partial \beta _{0}}\right] , \\
\tau _{\pi V}^{r}& =\frac{1}{5\gamma _{r}^{\pi }}\left[ r m^{2}\gamma
_{r-1}^{V}+m^{2}\beta _{0}\frac{\partial \gamma _{r-1}^{V}}{\partial \beta
_{0}}\right.  \notag \\
& \left. -\left( r+5\right) \gamma _{r+1}^{V}-\beta _{0}\frac{\partial
\gamma _{r+1}^{V}}{\partial \beta _{0}}\right] , \\
\ell_{\pi W}^{r}& =-\frac{1}{5\gamma _{r}^{\pi }}\left( m^{2}\gamma
_{r-1}^{W}-\gamma _{r+1}^{W}\right) , \\
\ell_{\pi V}^{r}& =-\frac{1}{5\gamma _{r}^{\pi }}\left( m^{2}\gamma
_{r-1}^{V}-\gamma _{r+1}^{V}\right) , \\
\lambda _{\pi W}^{r}& =\frac{1}{5\gamma _{r}^{\pi }}\left[ m^{2}\left( \frac{%
\partial \gamma _{r-1}^{W}}{\partial \alpha _{0}}+h_{0}^{-1}\frac{\partial
\gamma _{r-1}^{W}}{\partial \beta _{0}}\right) \right.  \notag \\
& -\left. \left( \frac{\partial \gamma _{r+1}^{W}}{\partial \alpha _{0}}%
+h_{0}^{-1}\frac{\partial \gamma _{r+1}^{W}}{\partial \beta _{0}}\right) %
\right] , \\
\lambda _{\pi V}^{r}& =\frac{1}{5\gamma _{r}^{\pi }}\left[ m^{2}\left( \frac{%
\partial \gamma _{r-1}^{V}}{\partial \alpha _{0}}+h_{0}^{-1}\frac{\partial
\gamma _{r-1}^{V}}{\partial \beta _{0}}\right) \right.  \notag \\
& \left. -\left( \frac{\partial \gamma _{\left( r+1\right) }^{V}}{\partial
\alpha _{0}}+h_{0}^{-1}\frac{\partial \gamma _{\left( r+1\right) }^{V}}{%
\partial \beta _{0}}\right) \right] , \\
\lambda _{\pi \pi }^{r}& =-\frac{1}{7\gamma _{r}^{\pi }}\left[ \left(
2r-2\right) m^{2}\gamma _{r-2}^{\pi }-\left( 2r+5\right) \gamma _{r}^{\pi }%
\right] , \\
\delta _{\pi \pi }^{r}& =\frac{n_{0}D_{20}^{-1}}{2\gamma _{r}^{\pi }}\left[
\left( J_{20}\frac{\partial \gamma _{r}^{\pi }}{\partial \alpha _{0}}+J_{10}%
\frac{\partial \gamma _{r}^{\pi }}{\partial \beta _{0}}\right) h_{0}\right. 
\notag \\
& \left. -J_{30}\frac{\partial \gamma _{r}^{\pi }}{\partial \alpha _{0}}%
-J_{20}\frac{\partial \gamma _{r}^{\pi }}{\partial \beta _{0}}\right]  \notag
\\
& -\frac{1}{6\gamma _{r}^{\pi }}\left[ \left( r-1\right) m^{2}\gamma
_{r-2}^{\pi }-\left( r+4\right) \gamma _{r}^{\pi }\right] .  \label{delta_pi}
\end{align}

In the above relaxation equations we have expressed the proper-time
derivative and spatial derivative of the coefficients from Eqs. (\ref%
{rho_scalar_final},\ \ref{rho_vector_final}, \ref{rho_tensor_final}) using
Eqs. (\ref{alpha_dot}, \ref{beta_dot}, \ref{u_dot}). Therefore, for any
coefficient $\gamma _{r}^{\Pi }$, $\gamma _{r}^{V}$, $\gamma _{r}^{W}$, or $%
\gamma _{r}^{\pi }$, collectively denoted by $\psi \left( \alpha _{0},\beta
_{0}\right) $, we used the following formula for the proper time derivative 
\begin{align}
\dot{\psi}& =\frac{n_{0}}{D_{20}}\left[ \left( J_{20}\frac{\partial \psi }{%
\partial \alpha _{0}}+J_{10}\frac{\partial \psi }{\partial \beta _{0}}%
\right) h_{0}\right.  \notag \\
& \left. -\left( J_{30}\frac{\partial \psi }{\partial \alpha _{0}}+J_{20}%
\frac{\partial \psi }{\partial \beta _{0}}\right) \right] \theta ,
\end{align}%
while for the spatial gradient of $\psi $ we used 
\begin{equation}
\nabla ^{\mu }\psi =\left( \frac{\partial \psi }{\partial \alpha _{0}}%
+h_{0}^{-1}\frac{\partial \psi }{\partial \beta _{0}}\right) \nabla ^{\mu
}\alpha _{0} - \beta _{0}\frac{\partial \psi }{\partial \beta _{0}} \dot{u}%
^{\mu }.
\end{equation}
Note that these two equations follow from the equations of ideal fluid
dynamics and we neglected terms from the general conservation equations
proportional to the dissipative fields or their derivatives.

\section{Choice of moment and coefficients in the massless limit}

\label{choice_of_moment}

As was shown in the previous section, \textit{once} the 14-moment
approximation is applied, \textit{any} moment of the Boltzmann equation will
lead to a closed set of equations which looks formally the same, see Eqs.\ (%
\ref{relax_bulk}), (\ref{relax_heat}), and (\ref{relax_stress}). This is
immediately apparent in these equations by the explicit dependence of the
transport coefficients on the index $r$. Barring any miraculous
cancellation, already at this point it is obvious that their values will in
general be different for different $r$. Thus, the 14-moment approximation
leads to an ambiguity, since it gives rise to an infinite set of equations
to describe a finite set of macroscopic variables.

In order to make this clear, we shall consider two different choices for the
moments of the Boltzmann equation: the first one is the traditional choice
of Israel and Stewart \cite{IS} and the other one was recently proposed by
Denicol, Koide, and Rischke (DKR) \cite{Denicol:2010xn}, following Grad's
original idea.

\subsection{Equations of motion of Israel and Stewart}

Israel and Stewart assumed that the equations of motion for the dissipative
currents could be extracted from the second moment of the Boltzmann equation 
\cite{Stewart_Review,Stewart_Proc,IS}, 
\begin{equation}
\partial _{\mu }\left\langle k^{\mu }k^{\nu }k^{\lambda }\right\rangle =\int
dKk^{\nu }k^{\lambda }C\left[ f\right] ,
\end{equation}%
with the equations for $\Pi $, $V^{\mu }$ or $W^{\mu }$, and $\pi ^{\mu \nu
} $ obtained using the following projections of the above equation, 
\begin{eqnarray}
u_{\nu }u_{\lambda }\partial _{\mu }\left\langle k^{\mu }k^{\nu }k^{\lambda
}\right\rangle &=&u_{\nu }u_{\lambda }\int dKk^{\nu }k^{\lambda }C\left[ f%
\right] ,  \label{IS_eq_scalar} \\
\Delta _{\lambda }^{\alpha }u_{\nu }\partial _{\mu }\left\langle k^{\mu
}k^{\nu }k^{\lambda }\right\rangle &=&\Delta _{\lambda }^{\alpha }u_{\nu
}\int dKk^{\nu }k^{\lambda }C\left[ f\right] ,  \label{IS_eq_vector} \\
\Delta _{\nu \lambda }^{\alpha \beta }\partial _{\mu }\left\langle k^{\mu
}k^{\nu }k^{\lambda }\right\rangle &=&\Delta _{\nu \lambda }^{\alpha \beta
}\int dKk^{\nu }k^{\lambda }C\left[ f\right] ,  \label{IS_eq_tensor}
\end{eqnarray}
respectively, \textit{together} with the 14-moment approximation.

As a matter of fact, Eqs. (\ref{IS_eq_scalar}), (\ref{IS_eq_vector}) and (%
\ref{IS_eq_tensor}) can be identified as the equations for $\rho _{3}$, $%
\rho _{2}^{\mu }$, and $\rho _{1}^{\mu \nu }$. These equations have already
been calculated with the 14-moment approximation, Eqs.\ (\ref{relax_bulk}), (%
\ref{relax_heat}), and (\ref{relax_stress}). Thus, using the indices $r=3$
(for the scalar moment), $r=2$ (for the vector), and $r=1$ (for the
irreducible second-rank tensor), we obtain the IS equations.

Even though this choice of moments was never clearly justified, this
prescription is widely employed in relativistic kinetic theory. However, it
was recently found that, at least for some cases, the IS equations are not
in good agreement with the numerical solution of the Boltzmann equation \cite%
{Huovinen:2008te,Molnar:2009pq,El:2009vj,Bouras:2009nn,Bouras:2010nt,Bouras:2010hm}%
. Also, the transport coefficients obtained by Israel and Stewart do not
coincide with quantum-field theoretical calculations \cite{Denicol:2010br}.

\subsection{Equations of motion directly from the dissipative currents}

In the second choice, the equations of motion for the dissipative currents
are obtained from the moments $\rho _{0}$, $\rho _{0}^{\mu }$, and $\rho
_{0}^{\mu \nu }$ which exactly correspond to the dissipative currents, see
Eqs.\ (\ref{rho_0}), (\ref{rho_mu_0}), and (\ref{rho_mu_nu_0}). Here, we
have already fixed the LRF in accordance with the Landau picture. Then, the
equations of motion for the dissipative currents emerge directly from their
definitions as 
\begin{eqnarray}
\dot{\Pi} &=&-\frac{1}{3}m^{2}\int dK\delta \dot{f}_{\mathbf{k}},
\label{Exact1} \\
\dot{V}^{\left\langle \mu \right\rangle } &=&\int dKk^{\left\langle \mu
\right\rangle }\delta \dot{f}_{\mathbf{k}},  \label{Exact2} \\
\dot{\pi}^{\left\langle \mu \nu \right\rangle } &=&\int dKk^{\left\langle
\mu \right. }k^{\left. \nu \right\rangle }\delta \dot{f}_{\mathbf{k}}.
\label{Exact3}
\end{eqnarray}%
As already mentioned, these equations are related to the equations for $\rho
_{0}=-3\Pi /m^{2}$, $\rho _{0}^{\mu }=V^{\mu }$, and $\rho _{0}^{\mu \nu
}=\pi ^{\mu \nu }$, that is, Eqs.\ (\ref{scalar_rho}), (\ref{vector_rho}),
and (\ref{tensor_rho}), for $r=0$. In this scenario, the fluid-dynamical
equations (\ref{bla1},\ref{bla2}) are closed by Eqs.\ (\ref{relax_bulk}), (%
\ref{relax_heat}), and (\ref{relax_stress}), with $r=0$, which correspond to
Eqs.\ (\ref{Exact1}), (\ref{Exact2}), and (\ref{Exact3}), once the 14-moment
approximation is applied.

It was shown that this method can successfully reproduce the numerical
solution of the Boltzmann equation for the simple one-dimensional scaling
expansion \cite{Denicol:2010xn}. It is also important to mention that the
transport coefficients of this kinetic calculation are consistent with those
calculated from quantum field theory with the method proposed in Ref.\ \cite%
{Denicol:2010br}.

\begin{table*}[tbp]
\begin{center}
\begin{tabular}{|c|c|c|c|c|c|c|c|c|c|c|}
\hline
& $\eta^{r} $ & $\tau^{r} _{\pi }$ & $\lambda^{r} _{\pi \pi }$ & $%
\lambda^{r} _{\pi V}$ & $\lambda^{r} _{\pi W}$ & $\delta^{r} _{\pi \pi }$ & $%
\ell^{r} _{\pi V}$ & $\ell^{r} _{\pi W}$ & $\tau^{r} _{\pi V}$ & $\tau^{r}
_{\pi W}$ \\ \hline\hline
$r=1$ \,(IS) & ${6}\sigma _{T}^{-1}/({5\beta _{0}})$ & ${9}\sigma _{T}^{-1}/(%
{5}n_{0})$ & $1$ & ${-1}/({3\beta _{0}})$ & $1/12$ & $2/3$ & ${-2}/({3\beta
_{0}})$ & $1/3$ & ${8}/({3\beta _{0}})$ & $-5/3$ \\ 
$r=0$ \,(DKR) & ${4}\sigma _{T}^{-1}/({3\beta _{0}})$ & ${5}\sigma
_{T}^{-1}/({3}n_{0})$ & $5/7$ & $0$ & 0 & $2/3$ & $0$ & $1/5$ & $0$ & $-1$
\\ \hline
\end{tabular}%
\end{center}
\caption{{\protect\small The coefficients for the shear-stress tensor in the
two approaches for the classical gas with constant cross section in the
ultrarelativistic limit.}}
\label{shear_massless}
\end{table*}
\begin{table*}[tbp]
\begin{center}
\begin{tabular}{|c||c|c|c|c|c|c|c|c|}
\hline
& $\kappa^{r} _{q}$ & $\tau^{r} _{V}$ & $\psi^{r} _{W}$ & $\tau^{r} _{W}$ & $%
\delta^{r} _{VV}$ & $\delta^{r} _{WW}$ & $\lambda^{r} _{VV}$ & $\lambda^{r}
_{WW}$ \\ \hline\hline
$r=2$ \,(IS) & ${2\sigma }_{T}^{-1}$ & ${5}\sigma _{T}^{-1}/(2n_{0})$ & $-{%
\beta _{0}}/{4}$ & $5\sigma _{T}^{-1}/(2n_{0})$ & $-1$ & ${-4}/{3}$ & $-7/5$
& ${9}/5$ \\ 
$r=0$ \,(DKR) & ${3}\sigma _{T}^{-1}$ & $9\sigma _{T}^{-1}/(4n_{0})$ & $-{%
\beta _{0}}/{4}$ & $9\sigma _{T}^{-1}/(4n_{0})$ & $-1$ & $-4/{3}$ & $-3/5$ & 
$1$ \\ \hline
\end{tabular}%
\end{center}
\caption{{\protect\small The coefficients for the particle and energy
diffusion in the two approaches for a classical gas with constant cross
section in the ultrarelativistic limit.}}
\label{diff_massless}
\end{table*}

\begin{table*}[tbp]
\begin{center}
\begin{tabular}{|c|c|c|c|c|c|c|}
\hline
& $\lambda^{r} _{V \pi}$ & $\lambda^{r} _{W \pi}$ & $\ell^{r}_{V \pi}$ & $%
\ell^{r}_{W \pi}$ & $\tau^{r} _{V \pi}$ & $\tau^{r} _{W \pi}$ \\ \hline\hline
$r=2$ \,(IS) & ${-3\beta }_{0}/40$ & $3/10$ & $-\beta _{0}/20$ & ${1}/{5}$ & 
0 & ${0}$ \\ 
$r=0$ \,(DKR) & $-\beta _{0}/20$ & $1/5$ & $-\beta _{0}/20$ & $1/{5}$ & 0 & $%
0$ \\ \hline
\end{tabular}%
\end{center}
\caption{{\protect\small The coefficients which couple shear stress and
particle or energy diffusion in the two approaches for a classical gas with
constant cross section in the ultrarelativistic limit.}}
\label{diff_massless2}
\end{table*}

\subsection{Comparison of choices}

\begin{figure}[tbp]
\includegraphics[scale=1.0]{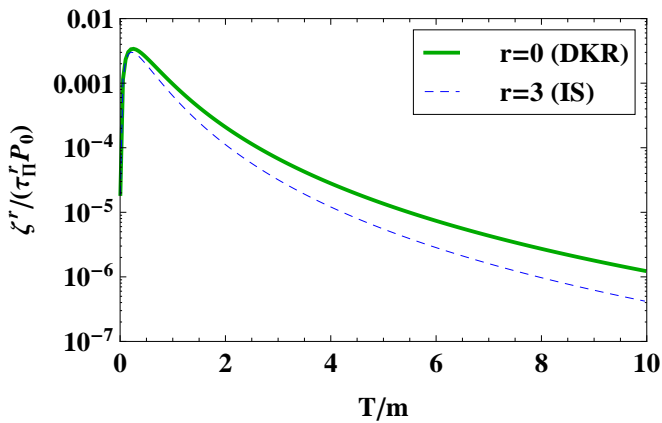}
\caption{The coefficient $\protect\beta _{\Pi }^{r}$ normalized by the
pressure $P_{0}$. The cases $r=3$ (dashed line) and $r=0$ (solid) line
correspond to the choices by Israel and Stewart and by Denicol, Koide and
Rischke.}
\label{Bulk_Coef}
\end{figure}

\begin{figure}[tbp]
\includegraphics[scale=1.0]{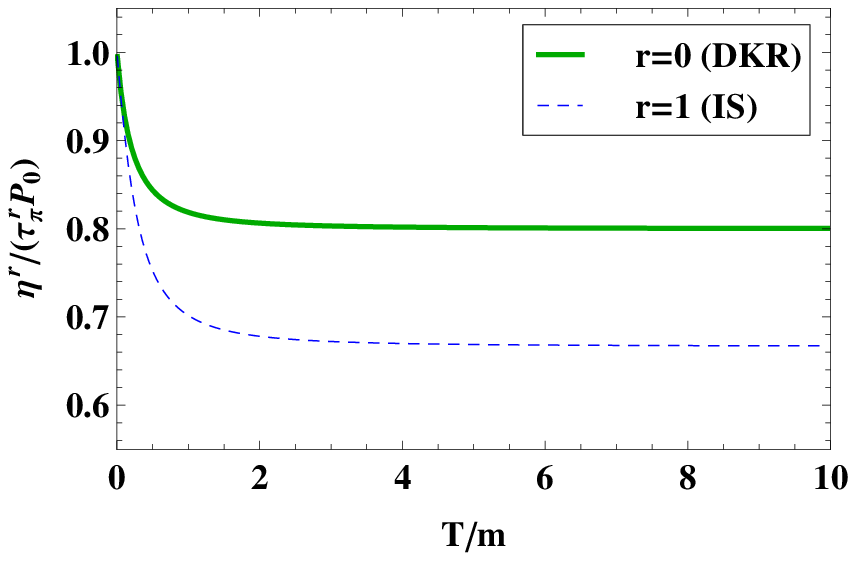}
\caption{The coefficient $\protect\beta _{\protect\pi }^{r}$ normalized by
the pressure $P_{0}$. The cases $r=1$ (dashed line) and $r=0$ (solid) line
correspond to the choices by Israel and Stewart and by Denicol, Koide and
Rischke.}
\label{Shear_Coef}
\end{figure}

\begin{figure}[tbp]
\includegraphics[scale=1.0]{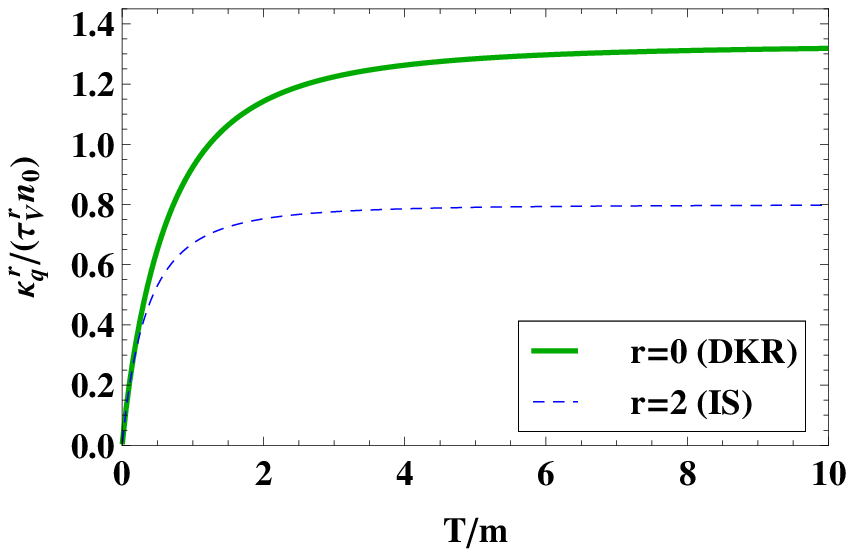}
\caption{The coefficient $\protect\beta _{V}^{r}$ normalized by the pressure 
$P_{0}$. The cases $r=2$ (dashed line) and $r=0$ (solid) line correspond to
the choices by Israel and Stewart and by Denicol, Koide and Rischke.}
\label{Part_Coef}
\end{figure}

In order to understand the difference between the two approaches discussed
above, we calculate the coefficients $\beta _{\Pi }^{r}=-\zeta ^{r}/\tau
_{\Pi }^{r}$, $\beta _{\pi }^{r}=\eta ^{r}/\tau _{\pi }^{r}$, and $\beta
_{V}^{r}=\kappa _{q}^{r}/\left( \tau _{V}^{r}\beta _{0}^{2}h_{0}^{2}\right) $%
. These coefficients, normalized by the pressure or particle density, are
shown in Figs.\ \ref{Bulk_Coef}, \ref{Shear_Coef}, and \ref{Part_Coef},
respectively. The calculations were done for a classical gas with fixed
chemical potential, $\mu =0$.

We see that both calculations converge at low temperatures but deviate
considerably at high temperatures. This behavior should be qualitatively the
same for any choice of moment because all irreducible moments of the same
rank converge to the same values in the non-relativistic limit (multiplied
by a different power of the mass). Thus, differences between the choice of
moment will only appear in the relativistic limit.

The coefficients in the ultrarelativistic limit, $m/T=0$, for a classical
gas with constant cross section, can be calculated analytically. These are
collected for the shear viscosity and particle diffusion in Tables \ref%
{shear_massless}, \ref{diff_massless}, and \ref{diff_massless2}. Note that,
in this limit, the bulk viscous pressure vanishes and was not considered.
For the relaxation times, $\tau _{V}$ and $\tau _{\pi }$, and transport
coefficients, $\eta $ and $\kappa _{V}$, the differences are of the order of 
$10-20$~\%, but for other coefficients the differences can be more
significant.

The coefficients $\ \lambda _{q\pi}^{r}, \ell_{q\pi}^{r}$, and $\tau
_{q\pi}^{r}$ in Eq.\ (\ref{relax_heat}) are calculated both in the Eckart
and Landau frames, see Table 3. For example, in the Landau frame we only
have an equation for $V^{\mu }$ and hence $\lambda _{V\pi}^{r}=\lambda
_{q\pi}^{r}$ is given exactly by Eq.\ (\ref{lambda_pi_q}), while in the
Eckart frame we only obtain an equation for $W^{\mu }$ so that, $\lambda
_{\pi W}^{r}=$ $\lambda _{\pi q}^{r}/\psi _{W}^{r}$. However, if we use the
definition of the heat flow in either frame, i.e., $q^{\mu} = -h_0 V^{\mu}$
or $q^{\mu} = W^{\mu}$, then these coefficients lead to the same values for
a classical gas where $\psi _{r}^{W} =-h_0^{-1}$.

\section{Conclusions}

\label{conclusions}

In this work we have reviewed the 14-moment approximation proposed by Israel
and Stewart and discussed the ambiguities of this approach. We started by
introducing a general expansion of the single-particle distribution function
in terms of its moments. For this purpose, we constructed an orthonormal
expansion basis which allowed us to establish exact relations between the
expansion parameters and the moments of the distribution function. We then
proceeded to derive the exact equations of motion for these moments.

Next, we showed how the 14-moment approximation can be obtained as a
truncation of this general expansion of the distribution function. We proved
that, once the 14-moment approximation has been applied, it is possible to
derive an infinite number of fluid-dynamical equations, all having the same
general structure but with different transport coefficients. This means that
the 14-moment approximation is not able to provide a unique theory of fluid
dynamics and, in this sense, is ambiguous. In Sec.~\ref{choice_of_moment} we
analysed two different choices for the moment equations: the one
corresponding to Israel and Stewart \cite{IS}, and the other one to that of
Denicol, Koide, and Rischke~\cite{Denicol:2010xn}. It is also worth to
mention that in this derivation we obtained terms that were neglected in the
original work of IS \cite{IS}, as was already presented in Ref.\ \cite%
{Betz:2008me,Betz:2010cx}.

We also remark that the solutions of the IS equations were already compared
to the numerical solutions of the Boltzmann equation for the so-called
Bjorken-scaling problem in Refs.\ \cite%
{Huovinen:2008te,Molnar:2009pq,El:2009vj} and for the relativistic Riemann
problem in Ref.\ \cite{Bouras:2009nn,Bouras:2010nt,Bouras:2010hm}. It was
demonstrated that IS theory is in relatively good agreement with the
numerical solutions of the Boltzmann equation only if the Knudsen number is
sufficiently small. Note that these comparisons did not include all
non-linear terms and transport coefficients derived in this work. On the
other hand, in Ref.~\cite{Denicol:2010xn} it was shown that, in contrast to
IS theory, the direct method gives a much better agreement with the
numerical solution of the Boltzmann equation up to very large Knudsen
numbers.

Before closing we mention that recently the me\-thod presented in this work
was extended to include $14+9\times n$ moments. It was explicitly shown how
to successively improve the expression for the transport coefficients by
extending the number of moments from $n=0$ to $n=1,2$, and $3$ \cite%
{Denicol:2012cn}. Furthermore, it was also shown that the equations of
motion can be closed in terms of 14 dynamical variables without making use
of the direct truncation of the moment expansion, the 14-moment
approximation. This was obtained by a separation of the microscopic time
scales and a power-counting scheme in Knudsen and inverse Reynolds number.
The equations of motion can be closed in terms of only 14 dynamical
variables, as long as we only keep terms of second order in Knudsen and/or
inverse Reynolds number.

\begin{acknowledgement}
This work was supported by the Helmholtz International Center for FAIR within the
framework of the LOEWE program launched by the State of Hesse. 
The work of H.N.\ was supported by the Extreme Matter Institute (EMMI)
and the Aca\-de\-my of Finland, Project No. 133005, that
of E.M.\ by the Hungarian National Development Agency OTKA/NF\"{U} 81655.
\end{acknowledgement}

\appendix 

\section{The collision integral in the massless limit}

\label{relax_times}

In this Appendix, we calculate the collision tensor defined in Eq.\ (\ref%
{X_4rank_tens_def}). For a classical gas with constant cross section, 
\begin{align}
X_{r}^{\mu \nu \alpha \beta }& =\frac{1}{\nu }\int dKdK^{\prime
}dPdP^{\prime }W_{\mathbf{kk}\prime \rightarrow \mathbf{pp}\prime }f_{0%
\mathbf{k}}f_{0\mathbf{k}\prime }  \notag \\
& \times E_{\mathbf{k}}^{r}k^{\mu }k^{\nu }\left( p^{\alpha }p^{\beta
}+p^{\prime \alpha }p^{\prime \beta }-k^{\alpha }k^{\beta }-k^{\prime \alpha
}k^{\prime \beta }\right) .  \label{X_4_classical}
\end{align}%
First we define the total cross section as 
\begin{equation}
\sigma _{T}\left( s\right) =\frac{1}{\nu }\int 2\pi \sin \Theta _{s}d\Theta
_{s}\,\sigma \left( s,\Theta _{s}\right) ,
\end{equation}%
where $\sigma \left( s,\Theta _{s}\right) $ is the differential cross
section, $s$ is a collision invariant, i.e., a Mandelstam variable, and $%
\Theta _{s}$ is the scattering angle, 
\begin{eqnarray}
s &\equiv &\left( k^{\mu }+k^{\prime \mu }\right) ^{2}=\left( p^{\mu
}+p^{\prime \mu }\right) ^{2}, \\
\Theta _{s} &=&\arccos \left[ \frac{\left( k^{\mu }-k^{\prime \mu }\right)
\left( p_{\mu }-p_{\mu }^{\prime }\right) }{\left( k^{\mu }-k^{\prime \mu
}\right) ^{2}}\right] .
\end{eqnarray}%
The transition rate $W_{\mathbf{kk}\prime \rightarrow \mathbf{pp}\prime }$
is written in terms of the differential cross section as 
\begin{equation}
W_{\mathbf{kk}\prime \rightarrow \mathbf{pp}\prime }=\left( 2\pi \right)
^{6}s\,\sigma \left( s,\Theta _{s}\right) \delta ^{4}\left( k^{\mu
}+k^{\prime \mu }-p^{\mu }-p^{\prime \mu }\right) .  \label{transition_rate}
\end{equation}

In order to simplify the calculations we divide$\ X_{r}^{\mu \nu \alpha
\beta }$\ into gain and loss parts,%
\begin{equation}
X_{r}^{\mu \nu \alpha \beta }= \mathcal{G}_{r}^{\mu \nu \alpha \beta }- 
\mathcal{L}_{r}^{\mu \nu \alpha \beta },
\end{equation}%
where%
\begin{align}
\mathcal{G}_{r}^{\mu \nu \alpha \beta }& =\frac{1}{\nu }\int dKdK^{\prime
}dPdP^{\prime }W_{\mathbf{kk}\prime \rightarrow \mathbf{pp}\prime }f_{0%
\mathbf{k}}f_{0\mathbf{k}^{\prime }}  \notag \\
& \times E_{\mathbf{k}}^{r}k^{\mu }k^{\nu }\left( p^{\alpha }p^{\beta
}+p^{\prime \alpha }p^{\prime \beta }\right) , \\
\mathcal{L}_{r}^{\mu \nu \alpha \beta }& =\frac{1}{\nu }\int dKdK^{\prime
}dPdP^{\prime }W_{\mathbf{kk}\prime \rightarrow \mathbf{pp}\prime }f_{0%
\mathbf{k}}f_{0\mathbf{k}^{\prime }}  \notag \\
& \times E_{\mathbf{k}}^{r}k^{\mu }k^{\nu } \left( k^{\alpha }k^{\beta
}+k^{\prime \alpha }k^{\prime \beta }\right).
\end{align}%
The tensor $\mathcal{L}_{r}^{\mu \nu \alpha \beta }$ can be directly
integrated and written in terms of the total cross section, 
\begin{align}
\mathcal{L}_{r}^{\mu \nu \alpha \beta }& =\frac{1}{2}\int dKdK^{\prime }f_{0%
\mathbf{k}}f_{0\mathbf{k}^{\prime }} \, \sigma _{T}\left( s\right) \sqrt{%
s\left( s-4m^{2}\right) }  \notag \\
& \times E_{\mathbf{k}}^{r}k^{\mu }k^{\nu }\left( k^{\alpha }k^{\beta
}+k^{\prime \alpha }k^{\prime \beta }\right) ,
\end{align}%
where we used 
\begin{eqnarray}
&&\frac{1}{\nu } \int \frac{d^{3}\mathbf{p}}{p^{0}}\frac{d^{3}\mathbf{p}%
^{\prime }}{p^{\prime 0}} \, s\, \sigma \left( s,\Theta _{s}\right) \delta
_{kk^{\prime }\rightarrow pp^{\prime }}^{4}  \notag \\
&=&\frac{1}{2} \sigma _{T}\left( s\right) \sqrt{s\left( s-4m^{2}\right) } .
\end{eqnarray}%
For the tensor $\mathcal{G}_{r}^{\mu \nu \alpha \beta }$ we first introduce
the total momentum and corresponding projection orthogonal to it,%
\begin{eqnarray}
P_{T}^{\mu } &=&k^{\mu }+k^{\prime \mu }=p^{\mu }+p^{\prime \mu }, \\
\Delta _{P_T}^{\mu \nu } &=&g^{\mu \nu }-\frac{P_{T}^{\mu }P_{T}^{\nu }}{s}.
\end{eqnarray}%
Now, the $p-$dependent part of the integral can be written as,%
\begin{eqnarray}
&&\frac{1}{\nu }\int \frac{d^{3}\mathbf{p}}{p^{0}}\frac{d^{3}\mathbf{p}%
^{\prime }}{p^{\prime 0}} \, s \, \sigma \left( s,\Theta _{s}\right)
p^{\alpha }p^{\beta }\delta _{kk^{\prime }\rightarrow pp^{\prime }}^{4} 
\notag \\
&=&B_{1} P_{T}^{\alpha }P_{T}^{\beta } +B_{2}\Delta _{P_T}^{\alpha \beta },
\end{eqnarray}%
where in the center-of-momentum frame in which $P_{T}^{\mu }=\left( \sqrt{s}%
,0,0,0\right) $ and $\Delta _{P_T}^{\mu \nu }p_{\mu }p_{\nu }=-\left\vert 
\mathbf{p}\right\vert ^{2}$, we obtain 
\begin{eqnarray}
B_{1} &=&\frac{1}{4}\sigma _{T}\left( s\right) \sqrt{s\left( s-4m^{2}\right) 
}, \\
B_{2} &=&-\frac{\sqrt{s}}{12}\sigma _{T}\left( s\right) \left( \sqrt{s-4m^{2}%
}\right) ^{3}.
\end{eqnarray}

In the massless limit, the above results simplify considerably, 
\begin{align}
\mathcal{L}_{r}^{\mu \nu \alpha \beta }& =\frac{1}{2}\int dKdK^{\prime }f_{0%
\mathbf{k}}f_{0\mathbf{k}^{\prime }}\, s \,\sigma _{T}\left( s\right)  \notag
\\
& \times E_{\mathbf{k}}^{r}k^{\mu }k^{\nu }\left( k^{\alpha }k^{\beta
}+k^{\prime \alpha }k^{\prime \beta }\right) , \\
\mathcal{G}_{r}^{\mu \nu \alpha \beta }& =\frac{1}{3}\int dKdK^{\prime }f_{0%
\mathbf{k}}f_{0\mathbf{k}^{\prime }}\, s \, \sigma _{T}\left( s\right) 
\notag \\
& E_{\mathbf{k}}^{r}k^{\mu }k^{\nu }\left( P_{T}^{\alpha }P_{T}^{\beta }-%
\frac{s}{4}g^{\alpha \beta }\right) .
\end{align}%
From here on, we will consider only the case of constant cross section.
Then, using $s\equiv 2\left( k^{\mu }k_{\mu }^{\prime }\right) =2\left(
p^{\mu }p_{\mu }^{\prime }\right) $, we directly obtain 
\begin{align}
\mathcal{L}_{r}^{\mu \nu \alpha \beta }& = \sigma _{T}\left\langle E_{%
\mathbf{k}}^{r}k^{\mu }k^{\nu }k^{\alpha }k^{\beta }k^{\kappa }\right\rangle
_{0}\left\langle k_{\kappa }\right\rangle _{0}  \notag \\
& +\sigma _{T}\left\langle E_{\mathbf{k}}^{r}k^{\mu }k^{\nu }k^{\kappa
}\right\rangle _{0}\left\langle k^{\alpha }k^{\beta }k_{\kappa
}\right\rangle _{0},
\end{align}%
and%
\begin{align}
\mathcal{G}_{r}^{\mu \nu \alpha \beta }& = \frac{2}{3}\sigma
_{T}\left\langle E_{\mathbf{k}}^{r}k^{\mu }k^{\nu }k^{\alpha }k^{\beta
}k^{\kappa }\right\rangle _{0}\left\langle k_{\kappa }\right\rangle _{0} 
\notag \\
& +\frac{4}{3}\sigma _{T}\left\langle E_{\mathbf{k}}^{r}k^{\mu }k^{\nu
}k^{\kappa }k^{\left( \alpha \right. }\right\rangle _{0}\left\langle
k^{\left. \beta \right) }k_{\kappa }\right\rangle _{0}  \notag \\
& +\frac{2}{3}\sigma _{T}\left\langle E_{\mathbf{k}}^{r}k^{\mu }k^{\nu
}k^{\kappa }\right\rangle _{0}\left\langle k^{\alpha }k^{\beta }k_{\kappa
}\right\rangle _{0}  \notag \\
& -\frac{1}{3}\sigma _{T}g^{\alpha \beta }\left\langle E_{\mathbf{k}%
}^{r}k^{\mu }k^{\nu }k^{\kappa }k^{\lambda }\right\rangle _{0}\left\langle
k_{\kappa }k_{\lambda }\right\rangle _{0},
\end{align}%
Finally, using the definition of the thermodynamic integrals from Eq.\ (\ref%
{Inq}) we obtain 
\begin{align}
X_{r}^{\mu \nu \alpha \beta }& =- \frac{\sigma _{T}}{3}\, I_{r+5}^{\mu \nu
\alpha \beta \kappa }I_{1,\kappa }+ \frac{4\sigma _{T}}{3}\, I_{r+4}^{\mu
\nu \kappa \left( \alpha \right. }I_{2,\kappa }^{\left. \beta \right) } 
\notag \\
& - \frac{\sigma _{T}}{3}\, I_{r+3}^{\mu \nu \kappa }I_{3,\kappa }^{\alpha
\beta }-\frac{\sigma _{T}}{3}\, g^{\alpha \beta }I_{r+4}^{\mu \nu \kappa
\lambda }I_{2,\kappa \lambda }.
\end{align}
Therefore, the different projections are given as 
\begin{align}
X_{r,1}& \equiv X_{r}^{\mu \nu \alpha \beta }u_{\mu }u_{\nu }u_{\alpha
}u_{\beta },  \notag \\
& =-\sigma _{T}\left[ \frac{1}{3}\left(
I_{r+5,0}I_{10}+I_{r+3,0}I_{30}\right) -8I_{r+4,1}I_{21}\right] , \\
X_{r,3}& \equiv \frac{1}{3}X_{r}^{\mu \nu \alpha \beta }u_{\mu }\Delta _{\nu
\alpha }u_{\beta }  \notag \\
& =\frac{\sigma _{T}}{3}\left[
I_{r+5,1}I_{10}-4I_{r+4,1}I_{21}-I_{r+3,1}I_{31}\right] ,
\end{align}%
and%
\begin{align}
X_{r,4}& \equiv X_{r}^{\mu \nu \alpha \beta }\Delta _{\mu \nu \alpha \beta }
\notag \\
& =-\frac{2\sigma _{T}}{3}\left[ I_{r+5,2}I_{10}+4I_{r+4,2}I_{21}\right] .
\end{align}

In order to calculate the coefficients in the massless Boltzmann limit, we
use the following formula for the thermodynamic integrals 
\begin{equation}
I_{n+r,q}\left( \alpha _{0},\beta _{0},m\rightarrow 0\right) =\frac{%
P_{0}\left( r+n+1\right) !}{2\beta _{0}^{r+n-2}(2q+1)!!},  \notag
\end{equation}%
where $P_{0}=ge^{\alpha _{0}}\beta _{0}^{-4}/\pi ^{2}$, hence%
\begin{align}
X_{r,1}& =-\frac{\sigma _{T}P_{0}^{2}\left( r+4\right) !}{6\beta _{0}^{r+2}}%
\left( r^{2}+3r+2\right) , \\
X_{r,3}& =\frac{\sigma _{T}P_{0}^{2}\left( r+4\right) !}{18\beta _{0}^{r+2}}%
\left( r^{2}+7r+6\right) , \\
X_{r,4}& =-\frac{\sigma _{T}P_{0}^{2}\left( r+5\right) !}{45\beta _{0}^{r+2}}%
\left( r+10\right) .
\end{align}


\end{document}